\documentclass{lmcs} 
\pdfoutput=1

\usepackage[utf8]{inputenc}
\usepackage[T1]{fontenc}
\usepackage{lastpage}
\lmcsdoi{16}{3}{15}
\lmcsheading{}{\pageref{LastPage}}{}{}%
{Dec.~07,~2018}{Sep.~02,~2020}{}


\usepackage{amssymb}
\usepackage[all]{xy}

\let\oldin\in
\renewcommand{\in}{{\oldin}}

\newcommand\noi\noindent
\newcommand\cour[1]{{\tt #1}}

\renewcommand\int{{\mathbb Z}}
\newcommand{\set}[1]{\{#1\}}
\newcommand\lr[1]{\langle #1\rangle}
\newcommand\UU{{\mathcal U}}
\newcommand{\T}{{\mathcal T}}
\newcommand\MM{{\mathcal M}}
\newcommand\be{{\beta\eta}}

\newcommand\eqdf{\mathbin{{\triangleq}}}
\newcommand\bdf{\begin{defi}}
\newcommand\edf{\end{defi}}
\newcommand\bdfC{\begin{defiC}}
\newcommand\edfC{\end{defiC}}
\newcommand\bprop{\begin{prop}}
\newcommand\eprop{\end{prop}}
\newcommand\bpropC{\begin{propC}}
\newcommand\epropC{\end{propC}}
\newcommand\bpf{\begin{proof}}
\newcommand\epf{\end{proof}}
\newcommand\bthm{\begin{thm}}
\newcommand\ethm{\end{thm}}
\newcommand\bthmC{\begin{thmC}}
\newcommand\ethmC{\end{thmC}}
\newcommand\bcor{\begin{cor}}
\newcommand\ecor{\end{cor}}
\newcommand\bcorC{\begin{corC}}
\newcommand\ecorC{\end{corC}}
\newcommand\bnot{\begin{nota}}
\newcommand\enot{\end{nota}}
\newcommand\bnnot{\begin{nota}}
\newcommand\ennot{\end{nota}}

\newcommand\brem{\begin{rem}}
\newcommand\erem{\end{rem}}
\newcommand\isb{=_\beta}
\newcommand\ninb{\notin_{\!\beta}}
\newcommand\Right{\Rightarrow}
\newcommand\hb[1]{{#1}}
\newcommand{\FV}[1]{\mathrm{FV}(#1)}
\newcommand\Lo{\Lambda^o}
\newcommand\eqb{=_{\beta}}
\newcommand\sbs\subseteq
\newcommand\imp{\;\Right\;}

\newcommand\subs{\subsection*}

\newcommand\bexc{\begin{exercise}}
\newcommand\eexc{\end{exercise}}

\newcommand\benum{\begin{enumerate}}
\newcommand\eenum{\end{enumerate}}

\newcommand\bex{\begin{exa}}
\newcommand\eex{\end{exa}}

\newcommand\barr{\begin{array}}
\newcommand\earr{\end{array}}

\newcommand\church{{\mathbf c}}
\newcommand\cnum[1]{{{\church}_{#1}}}

\newcommand\seM[1]{\sem{#1}_M}

\renewcommand\partial{\rightharpoondown}

\newcommand\nat{{\mathbb N} }
\newcommand\weg[1]{{}}
\newcommand\up{\uparrow}

\newcommand\ldana{[\![}
\newcommand\rdana{]\!]}
\newcommand\semantics[1]{{\ldana{#1}\rdana}}
\newcommand\sem[1]{\semantics{#1}}
\newcommand\beqn{\begin{eqnarray*}}
\newcommand\eeqn{\end{eqnarray*}}

\theoremstyle{definition}\newtheorem{exercise}[thm]{Exercise}
\theoremstyle{thmC}
\newtheorem{corC}[thm]{Corollary}
\theoremstyle{defC}
\newtheorem{pkC}[thm]{Problem/Koan}
\theoremstyle{plain}

\newcommand\bceqn{\begin{array}{rcll}}
\newcommand\eceqn{\end{array}}

\newcommand\redd[1][]{\mathrel{\rightarrow\mathrel{\mkern-14mu}\rightarrow}_{#1}}

\newcommand\red[1][]{\mathrel{\rightarrow}_{#1}}
\newcommand\redb{\mathrel{\red[\beta]}}
\newcommand\rede{\mathrel{\red[\eta]}}

\newcommand\vect[2]{#1_1,\ldots,#1_{#2}}
\newcommand\vecn[2]{#1_1\cdots #1_{#2}}

\newcommand\dder{\mathrel{\leftarrow\!\!\!\!\!\leftarrow}}

\newcommand\bsub{\begin{enumerate}}
\newcommand\esub{\end{enumerate}}
\newcommand\fit{\item}

\newcommand\btab{\begin{tabular}}
\newcommand\etab{\end{tabular}}

\newcommand\CCC{{\mathcal{C}}}
\newcommand\CC{{\mathcal{C}}}
\newcommand\bS{{\bf S}}
\newcommand\sS{{\sf S}}
\newcommand\cU{{\mathcal{U}}}
\newcommand\cA{{\mathcal{A}}}

\newcommand\bK{{\bf{K}}}

\newcommand{\lh}{\mathopen{\mbox{\hspace{0.05cm}\rule[0.8ex]{0.01ex}{1ex}
\kern-.35em{\rule[1.8ex]{0.6ex}{0.01ex}}}}}
\newcommand{\rh}{\mathclose{\mbox{\rule[1.8ex]{0.6ex}{0.01ex}
\kern-.35em{\rule[0.8ex]{0.01ex}{1ex}\hspace{0.05cm}}}}}
\newcommand{\num} [1] {\lh{#1}\rh}

\newcommand\sE{{\sf E}}
\newcommand\eval{\sE_0}
\newcommand\numm[1]{\num{#1}^m}
\newcommand\numn[1]{\num{#1}^m}
\newcommand\numb[1]{\num{#1}^{bb}}
\newcommand\R{{\sf R}}
\newcommand\Eb{\sE^{bb}}
\newcommand\var{{\tt var}}
\newcommand\app{{\tt app}}
\newcommand\abs{{\tt abs}}
\newcommand\appb{{\tt app}^{bb}}
\newcommand\absb{{\tt abs}^{bb}}
\newcommand\varb{{\tt var}^{bb}}

\newcommand\bC{{\bf C}}
\newcommand\C{{\bf C}}
\newcommand\U{{\sf U}}
\newcommand\I{{\sf I}}
\newcommand\K{{\sf K}}
\newcommand\Y{{\sf Y}}
\newcommand\B{{\sf B}}
\newcommand\om{{\omega}}
\newcommand\Om{{\Omega}}
\newcommand\und{\;\&\;}
\renewcommand\ddot{\mathbin{\tiny \ast}}
\newcommand\hoog[1]{\rule{0mm}{#1}}
\newcommand\hoogm{\hoog{1.1em}}

\newcommand\dwn{\\[.2em]}
\newcommand\Dwn{\\[1em]}

\newcommand\ul[1]{\underline{#1}}
\makeatletter 
  \newcommand\figcaption{\def\@captype{figure}\caption} 
  \newcommand\tabcaption{\def\@captype{table}\caption}
\makeatother
\newcommand\bfig{\begin{minipage}{\textwidth}\centering}
\newcommand\efig{\end{minipage}}

\newcommand\wit[1]{\rule{#1mm}{0mm}}
\newcommand\arr{\rightarrow}

\begin{document}

\title{Gems of Corrado B\"ohm}

\author[H.P.~Barendregt]{Henk Barendregt}	
\address{Faculty of Science,
Radboud University Nijmegen\\        
Box 9010\\
6500GL Nijmegen, The Netherlands}	
\email{henk@cs.ru.nl}  





\begin{abstract}  
  \noindent The main scientific heritage of Corrado B\"ohm consists of
  ideas about computing, concerning concrete algorithms, as well as
  models of computability. The following will be presented. 1.\ A
  compiler that can compile itself. 2.~Structured programming,
  eliminating the `goto' statement. 3.\ Functional programming and an
  early implementation. 4.\ Separability in
  $\lambda$-calculus. 5.\ Compiling combinators without parsing.
  6.\ Self-evaluation in $\lambda$-calculus.
\end{abstract}

\maketitle

\begin{center}
To the memory of Corrado B\"ohm (1923-2017)
\end{center}

\section*{Introduction}
As a tribute to Corrado B\"ohm this paper presents six brilliant
results of his and also discusses some of their later
developments. Most of the papers are written by B\"ohm with
co-authors. The result on elimination of the {\tt goto}, Section 2, is
written by Giuseppe Jacopini alone in the joint paper 
\cite{bohmjaco66}, but one may assume that B\"ohm as supervisor had
influenced the research involved, and therefore this result is
included here. This paper is written such that computer science
freshmen can read and understand it.

\section{Self-compilation}
In his PhD thesis \cite{bohm51} at the ETH Z\"urich, Corrado B\"ohm
constructed one of the first higher programming languages $L$ together
with a compiler for it. The compiler has the particular feature that
it is written in the language $L$ itself. This sounds like magic, but
it is not: if a programming language is capable of expressing any
computational process, then it should also be able to `understand
itself' (i.e.\ perform the computational task to translate it into
machine language). Later this property gave rise to `bootstrapping':
dramatically increasing efficiency and reliability of computer
programs, that seems as impossible as to pull oneself over a fence by
pulling one's bootstraps\footnote{In Europe the hyperbole for
  impossibility is the story of Baron (von) M\"unchhausen, who could
  get himself (and the horse on which he was seated) out of a swamp by
  pulling up his own hair.}. This gave rise to the term `booting a
computer'. The mechanism will be explained in this section.

\subsection{Algorithms, computers, and imperative programming}
An \emph{algorithm} is a recipe to compute an output from a given
input. Executing such a recipe basically consists in putting down
pebbles\footnote{The word `pebble' in Latin is `calculus'.} in a fixed
array of boxes and `replacing' these pebbles step by step. That is, a
pebble may be moved from one box to another one, be taken away, or new
ones may be added. Such a process is called a \emph{calculation} or
\emph{computation}.  As shown in \cite{turi37a}, all computational
tasks, like ``What is the square of 29?'', ``Put the following list of
words in alphabetical order'', or ``What does Wikipedia say about the
concept `bootstrap'?'', can be put in the format of shuffling pebbles
in boxes.

This view on computing holds for computations on an abacus, but also
for programmed \emph{computers}. A computer $M$ is a, usually
electronic, device with memory, that performs computations. The
pebbles are represented in this memory and the shuffling is done by
making stepwise changes. A simple conceptual computer is the Turing
Machine (TM). It consists of an infinite\footnote{Actual computers
  only have a finite amount of memory. Turing apparently didn't want
  to be technology dependent and conceived the Turing Machine with an
  idealized memory of infinitely many cells. But at any given moment
  in a computation only finitely many cells contain a 1.} tape
of discrete cells that can be numbered by the integers
$\int=\set{\cdots,-2,-1,0,1,2,\cdots}$. At every moment in the
computation only a finite number of these cells contain information,
either a 1 or nothing: the original TM was a 0-bit\footnote{In 0-bit
  machines counting happens in the $2^0$-ary, i.e.\ unary, system. In
  modern computers the cells are replaced by registers that contain a
  sequence of 64 or more bits that can be read or overwritten in
  parallel; moreover, the registers do not need to be looked up
  linearly, like on the tape of the TM, but there is fast access to
  each of them; one speaks of `random access memory' (RAM).} machine.
The machine can be in one of a finite number of \emph{states}. For a
computational problem the input, coded as a list of the symbols, is
written on the tape.  There is a read/write (R/W) head positioned on
one of the cells of the tape. Depending on the symbol $a$ that is
read, and the present state $s$, one of the following three actions is
performed: a (possibly different) symbol $a'$ is written on the cell
under the R/W-head, a (possibly different) state $s'$ is assumed, and
finally the head moves $\set{R,L,N}$ ($R$: one position to the right,
$L$: one position to the left, $N$: no moving). When finally no action
can be performed any longer, the resulting information on the tape
represents the output of the computation. Each Turing Machine is
determined by a finite table consisting of 5-tuples like
$\lr{a,s;a',s',\set{R,L,N}}$ that determine the changes.

Turing showed that there exists a particular kind of machine, called a
\emph{universal machine} $\cU$, that suffices to make arbitrary
computations.  Such a $\cU$ is conceptually easy. The set of 5-tuples
of a particular machine $\MM$ is presented as a table $T_\MM$ `in its
silicon'. A universal machine $\cU$ that imitates $\MM$, needs this
table $T_\MM$ as extra input in coded form, including the collection
of all states of $\MM$ (that may be more extensive than that of $\cU$)
and the present state of $\MM$, stored in a dedicated part of the
memory as the program (nowadays known as the `app') for $\MM$.  The
instruction table $T_\cU$ of $\cU$ stipulates that 1.\ it has to look
in $\T_\MM$ in order to see what is the present state of $\MM$, and to
know what to do next; and 2.\ to do this. The possibility of a
universal machine provides a model of computation in which a single
machine $\MM$, using programming language $M=L_\MM$, can perform any
computational job.  The nature of the actions of Turing machines,
described in their action tables, is rather imperative: overwrite
information, change state, move. For this reason the resulting
computational model is called \emph{imperative programming}.

In this paper we will consider a fixed universal machine $\MM$. Around
1950, when Corrado B\"ohm worked on his PhD, computers were
rare. Indeed, in 1954, in a country like the Netherlands there were
only three computers (at the Mathematical Center, the Royal
Meteorological Institute, and the National Phone Company) and no more
were deemed to be necessary!  Nowadays a standard car often has
on board in the order of $150$ (universal) computers in the form of
microprocessor chips.

A program in a given language $M$ for $\MM$ consists of a sequence of
\emph{statements} in $M$ that the machine `understands': it performs
intended changes on data represented in the memory of $\MM$. Such
programs are denoted by $p=p^M$, the optional superscript indicating
that the program is written in the language $M$.

\bdf \bsub\item There is a non-specified set $D$ (for data) consisting
of the intended objects on which computations take place.
\item 
The process of running program $p^M$ on input $x$ in $D$ is denoted by
$\set{p^M}(x)\footnote{Compound expressions like $\set{\set{c}(p)}(x)$
  make sense and will be used. But an expressions like
  $\set{q}(\set{p}(x))$ we will avoid, as one is forced to evaluate
  first the $\set{p}(x)$, which may be undefined; therefore even if
  $\forall y.\set{q}(y)\redd 0$, one doesn't always have
  $\set{q}(\set{p}(x))\redd 0$. See \cite[Exercise 9.5.13]{bare84} and
  \cite{bare75,bare96a}.}$.
If this process terminates with end
result $y$ (the output, again in $D$), then we
write $$\set{p^M}(x)\redd y.$$
\item It may be the case that $\set{p^M}(x)$ doesn't terminate. 
Then there is no output, and we write  $\set{p^M}(x)\!\up$.
\item The (operational) \emph{semantics} of $p^M$ is the partial
  map $\sem{p^M}\colon D\partial D$ defined as follows.  
\[\bceqn \sem{p^M}(x)&=&y,&\mbox{ if $\set{p^M}(x)\redd y$};\\
&=&\up,&\mbox{ if $\set{p^M}(x)\!\up$}.  \eceqn
\]
\esub\edf For $\set{\ }$ and $\sem{\ }$, that depend on $M$, we
sometimes write $\set{\ }_M$, $\sem{\ }_M$, respectively.

The difference between $\sem{p^M}(x)=y$ and $\set{p^M}(x)\redd y$ is
that the former is an identity, like $36^2=36\times 36$ that
holds by definition, whereas the latter requires a computation, like
$36\times 36\redd 1296$.  The sign `$\redd$' indicates that a
\emph{computation} has to be performed that takes time, consisting of
a sequence of a few or more steps that transform information.

\bprop\label{set.sem} If $\set{p^M}(x)$ terminates, then
$$\set{p^M}(x)\redd\sem{p^M}(x).$$
\eprop\bpf By definition.\epf

\subsection{Programming languages and compilers}
A human, having to write a correct and efficient program, better does
this in an understandable way, rather than in the form of recipes for
shuffling pebbles. One can use a \emph{programming language} $L$ for
this, in which computational tasks can be described more
intuitively. In \cite{bohm51} an early example of such a language $L$
is constructed.

\bdf \bsub\fit 
A programming language $L$ consists of programs $p$ that describe
computations according to (2).
\item $L$ comes with a (denotational) \emph{semantic function}
  $\sem{\ }_L\colon L\red (D\partial D)$. That is, to each $p^L\in L$
  it assigns a (possibly partial) function $\sem{p^L}_L\colon D\partial D$.
  \esub \edf

Technically speaking $M$ is also a programming language, the
\emph{machine language}, with its denotational semantics $\sem{-}_M$,
by definition equal to the operational one $\set{-}_M$. By contrast
other programming languages are called \emph{higher} programming
languages, that are intended to make the construction of programs more
easy. When one has a program $p^L$ described in a higher programming
language $L$ we want to have machine help from a universal machine to
obtain from input $x$ the output $\sem{p^L}_L(x)$.  We succeed if one
can translate $p^L$ in the `right way' into the machine language
$M$. This translating is called \emph{compiling}.

\bdf\label{comp.fct} A function $C\colon L_1\red L_2$, is called a
\emph{compiling function} if
$$\sem{C(p^{L_1})}_{L_2}=\sem{p^{L_1}}_{L_1}.$$ \edf In this paper, we
will usually consider only compilers into $L_2=M$. 

\bprop\label{comp.def-prop} If $C\colon L\red M$ is a compiling function, 
then \[\set{C(p^{L})}_M(x)\redd\sem{p^{L}}_L(x).\]\qedhere
\eprop\bpf 
One has by Proposition \ref{set.sem} and Definition \ref{comp.fct}
\[\set{C(p^L)}_M(x)\redd\sem{C(p^L)}_M(x)=\sem{p^L}_L(x).\qedhere\]
\epf

This shows that an intended computation using a $p^L\in L$, intended
to compute $\sem{p^L}_L(x)$, can in principle be replaced by a
computation using a $p^M\in M$, for which there is the support from the
machine $\MM$. We say: the computational task $\sem{p^L}_L(x)$ becomes
executable (by $\MM$).  In modern compilers the translation $L\red M$,
is often divided in literally hundreds of steps, using many
intermediate languages\footnote{For example one may have a long series
  of translations: $L\red L_1\red L_2\red \cdots \red L_n\red M.$}.
For example, the first step is the so called
\emph{lexing} that examines where every meaningful unit starts and
ends\footnote{Every student of a foreign language has to master this
  also: a stream of sounds `papafumeunepipe' has to be separated into words as follows
`papa fume une pipe'; only then one can translate further, into
  `father smokes a pipe'.}. At the end of the long translation process
one arrives at the language $M$. No need for further translation
occurs: in $\MM$ the programs in machine language are run by the
laws of physics (electrical engineering).

Compiling functions $C\colon L_1\red L_2$ are notably useful if the
translated program $C(p^{L_1})$ in $L_2$ in turn is
executable. Translating is a computational task and in principle
determining $C(p^L)$ can be done by hand. But since many programs,
also in a higher order programming language, may consist of several
million instructions, the computational task of compiling much
better be performed by a machine as well. A program that performs this
translation is called a \emph{compiler}. If such an automated
translation process is of any use, the compiler needs to be written
either in machine language $M$, or in another language $L$ for which
there is already another compiler from $L$ to $M$.

\bdf\label{def.compi} Let $C^{L_1}\colon L_1\red M$ be a compiling function. 
A \emph{compiler} for $C^{L_1}$ \emph{written in
language $L_2$} is a program $c^{L_1,L_2}$ such that
$$\sem{c^{L_1,L_2}}_{L_2}=C^{L_1}.$$
\edf

\weg{\bprop For a compiler $c^{L_1,L_2}$ for $L_1$,
written in $L_2$ by definition one has
$$\sem{c^{L_2}}_{L_2}(p^{L_1})=C(p^{L_1}).$$ \eprop} 

This is useful only if programs in $L_2$ are also executable.  
This is the case if $L_2=M$ or if there is already a compiler
from $L_2$ into $M$.
Two cases will be important in this paper. (1.)~$L_2=M$ and (2.)~$L_2=L_1$.

\subsection{Compilers written in machine language $M$}
First consider a compiler $c^L\colon L\red M$ written in machine
language $M$.

\bprop\label{compu.compi} 
Let $c^L\colon L\red M$ be a compiler for a compiling
function $C$. \bsub\item
For all programs $p^L$ written in $M$ one has $\set{c^L}_M(p^L)\redd C(p^L).$
\item A computational job $\sem{p^L}_L(x)$ can
be fully automated as follows.
$$\set{\set{c^L}(p^{L})}(x)\redd \set{C(p^L)}(x)\redd\sem{p^L}_L(x).$$
\esub\eprop\bpf\bsub\fit
By Definition \ref{def.compi} we have $\sem{c^L}_M=C$. 
Hence by Proposition \ref{set.sem} 
$$\set{c^L}(p^{L})\redd \sem{c^L}_M(p^L)=C(p^{L}).$$
\item It follows that
\[\bceqn
\set{\set{c^L}(p^{L})}(x)&\redd& \set{C(p^L)}(x),&\mbox{ by (1)},\\
&\redd&\sem{p^L}(x),&\mbox{ by Proposition \ref{comp.def-prop}}.
\null\hfill
\eceqn
\]
\par \vspace{-1.3\baselineskip}
\qedhere
\esub\epf

\bdf\label{cph} Let $c^L\colon L\red M$ be a compiler written in $M$.  
\bsub\item
By Proposition \ref{compu.compi}(2)
there are two computation phases towards  $\sem{p^L}(x)$:
$$\set{\set{c^L}(p^L)}(x)\redd^{{1}}
\set{C(p^L)}(x)\redd^{{2}}\sem{p^L}(x).$$ 
The first computation {{1}}, that is $\set{c^L}(p^L)\redd C(p^L)$, 
takes place in a time interval that is called \emph{compile-time}; 
the second computation
{{2}}, that is $\set{C(p^L)}(x)\redd \sem{p^L}(x)$, 
takes place in a time-interval 
that is called \emph{run-time}.  
\item
If for
programs $p^L$ and inputs $x$ (that interest us) the run-time
$\set{C(p^L)}(x)\redd \sem{p^L}(x)$ is short (for our purposes),
then the compiler $c^L$ is said to \emph{produce efficient
  code}. Note that this pragmatic definition depends only on the
compiling function $C=\sem{c^L}$, and not on its program, the compiler
itself.
\item If for programs $p^L$ (that interest us) the compile-time is
  short (for our purposes), then the compiler is said to be
  \emph{fast}.  Note that this notion does depend on the compiler
  $c^L$, and not on the compiling function $C=\sem{c^L}$. \weg{I do not
    understand: more refined compiling functions may require less
    efficient programs. Take for example a compiling
    function $C_1$ which translates a numeric expression from decimal
    to binary numerals and a compiling function $C_2$ which does the
    same but also put the expression in polish notation. I think the
    compile time of $C_2$ cannot be shorter than the compile time of
    $C_1$, and this shows that the compile time depends on the
    compiling function.}\esub \edf

\bprop For a programming language $L$, in which every program $p^L$ is
a sequence of statements consisting of a computable step, there exists
a simple compiler $c_I^{L,M}\colon L\red M$ written in $M$ for a
compiling function $C_I^L$, mimicking the steps in $L$ as steps in
$M$. Such a compiler is called a (simple)
\emph{interpreter}.\eprop\bpf [Sketch] Let
$p^L=s_1;s_2;\ldots;s_n$. Define
$C_I^L(p^L)=I(s_1);I(s_2);\ldots;I(s_n),$ where $I(s)$ mimics the
statement $s$ by a (small) program in $M$.\epf For complex
computational problems using a large program both the compile-time and
run-time consume considerable amounts of time. Often these are
bottlenecks for the feasibility of executing a program.  Moreover,
interpreters usually produce less efficient code than compilers, for
reasons to be discussed next.

\subsection{Compilers written in higher programming languages}\label{1.4}
Now we consider the task of writing a compiler $c=c^{L,M}\colon L\red
M$.  A compiler more complex than a simple interpreter is able to
look at the input program $p^L$ \emph{in its totality} and can
`reflect' (act) on it, enabling optimizations for the run-time of the
resulting code $p^M$. Such a compiler improves
efficiency\footnote{Software engineering studies ways to develop new
  versions of programs and compilers, in order to improve time
  performance and also to correct bugs (errors).}, using the power and
flexibility of $L$. With the right effort a compiler can be developed
that produces efficient code, so that to use such a compiler the
run-time performance of the translated programs are optimized. This
doesn't apply to the compile-time if compiler $c$ is written in
$M$, for which it is hard to achieve optimizations.

In his PhD thesis (1951) of just 50 pages Corrado B\"ohm designed a
programming language $L$ and constructed a compiler $c_B=c_B^{L,L}$,
\emph{in $L$ itself}.  This later made \emph{bootstrapping} possible:
producing not only efficient programs, but also making the compilation
process itself efficient. We will explain how this is
achieved. Suppose one has a compiler $c_B^{L,L}\in L$ that produces
efficient code (efficiently running programs). Here `efficient' is
used in a non-technical intuitive sense. In order to run $c_B^{L,L}$
one needs a simple interpreter $c_I^{L,M}\colon L\red M$, written in
$M$.  Now we will describe three ways of computing $\sem{p^L}_L(x)$,
that is, finding the intended result that program $p^L\in L$
has acting on input $x$.

1.\  Computing $\sem{p^L}_L(x)$ using the simple interpreter
$c_I^{L,M}$:\hoogm \[\bceqn
\set{\set{c_I^{L,M}}(p^L)}(x)&\redd&\set{C_I^L(p^L)}(x),&
\mbox{by {1 of Definition \ref{cph}(1)},} \\
&\redd& \sem{p^L}_L(x),&\mbox{by 2 of Definition \ref{cph}(1)}.
\eceqn\] This has both inefficient compile-time and
run-time.

2.\ \hoogm Better efficiency using $c_B^{L,L}$, run by the interpreter. Define
$c_B^{L,M}=C_I^L(c_B^{L,L})$, the interpreter applied to the compiler 
written in $L$. This can be precompiled
$$c_B^{L,M}=C_I^L(c_B^{L,L})\dder\set{c_I^{L,M}}(c_B^{L,L}),$$ as the
code of $C_B^L$ in the sense that $\sem{c_B^{L,M}}_M=C_B^L$.  One now
has
\[\bceqn \set{\set{c_B^{L,M}}(p^L)}(x)
&=&\set{\set{C_I^L(c_B^{L,L})}(p^L)}(x),&\mbox{by definition,}\\
&\redd&\set{\sem{c_B^{L,L}}_L(p^L)}(x),&\mbox{Prop.\ \ref{comp.def-prop} applied to {$C_{I}(c_B^{L,L})$,}}\\
&=&\set{C_B^L(p^L)}(x),&\mbox{as $\sem{c_B^{L,L}}_L=C_B^L$ by definition,}\\
&\redd&\sem{p^L}_L(x),&\mbox{Prop.\ \ref{comp.def-prop} applied to $C_B^L(p^L)$.}\eceqn\] Computing
$c_B^{L,M}$ is a one time job and, as the result can be stored, it
doesn't count in measuring efficiency.  The first computation $\redd$
counts as the compile time of $c_B^{L,M}$. But it is also the run-time
of $c_I^{L,M}$ (with compiling function $C_I^L$) and doesn't need to
be efficient.  The second computation $\redd$ is the run time of
$c^L_B$ (with compiling function $C_B^L$) and was assumed to be
efficient.  Therefore this computation has an efficient run-time, but
not necessarily an efficient compile-time.

\hoogm 3. Best efficiency using $c_B^{L,L}$: define $c_{B'}^{L,M}=C_B^L(c_B^{L,L})$,
the compiler applied to itself. This can be precompiled as follows.
$$c_{B'}^{L,M}=C_B^L(c_B^{L,L})
\dder\set{c_B^{L,M}}(c_B^{L,L})\dder\set{\set{c_I^{L,M}}(c_B^{L,L})}(c_B^{L,L}),$$ 
just requiring a one time computation. Then again  $\sem{c_{B'}^{L,M}}_M=C_B^L$,
but now
\[\bceqn
\set{\set{c^{L,M}_{B'}}(p^L)}(x)&=&\set{\set{C_B^L(c_B^{L,L})}(p^L)}(x),& 
\mbox{by definition,}\\ 
&\redd&\set{\sem{c^{L,L}_B}_L(p^L)}(x),&\mbox{Prop.\ \ref{comp.def-prop} applied to $C_B^L(c^{L,L}_B)$,}\\ 
&=&\set{C_B^L(p^L)}(x),&\mbox{as $C_B^L=\sem{c_B^{L,L}}_L$ by definition,}\\ 
&\redd&\sem{p^L}(x),&\mbox{Prop.\ \ref{comp.def-prop} applied to $C_B^L(p^L)$,}  \eceqn\] 
with both efficient compile and run-time, as both codes have
been generated by $C_B^L$.\Dwn
\bfig \fbox{$$\wit{6}\xymatrix{
    &&c_I^{L,M}c_B^{L,L}c_B^{L,L}p^Lx\ar@{->>}[d]|-{1.2}\hoog{1.8em}\\ &&\sem{c_I^{L,M}}_M(c_B^{L,L})p^Lx\ar@{=}[d]|-{1.6}\\ &c_I^{L,M}c_B^{L,L}p^Lx\ar@{->>}[d]|-{1.2}&
    C_B^L(c_B^{L,L})p^Lx\ar@{->>}[d]|-{1.5}\\ &\sem{c_I^{L,M}}_M(c_B^{L,L})p^Lx\ar@{=}[d]|-{1.6}&
    \sem{c_B^{L,L}}_L(c_B^{L,L})p^Lx\ar@{=}[d]|-{1.6}\\ {c_I^{L,M}p^Lx}\ar@{->>}[d]|-{\rm
      slow\ compiling\ 1.2\hspace{5.1em}} & \ul{C_I^L(c_B^{L,L})}p^Lx
    \ar@{->>}[d]|-{\rm slow\ compiling\ 1.5\hspace{5.1em}}&
    \ul{C_B^L(c_B^{L,L})}p^Lx
    \ar@{->>}[dl]|-{\rm \hspace{5.0em}1.5\ efficient\ compiling}\\ \sem{c_I^{L,M}}_M(p^L)x\ar@{=}[d]|-{1.6}&
    \sem{c_B^{L,L}}_L(p^L)x\ar@{=}[d]|-{1.6}\\ C_I^L(p^L)x\ar@{->>}[d]|-{\rm
      slow\ running\ 1.5\hspace{4.6em}}&
    C_B^L(p^L)x\ar@{->>}[dl]|-{\rm \hspace{5.6em}1.5\ efficient\ running}\\ \sem{p^L}_L(x)&&&\raisebox{-1.7em}{\ }
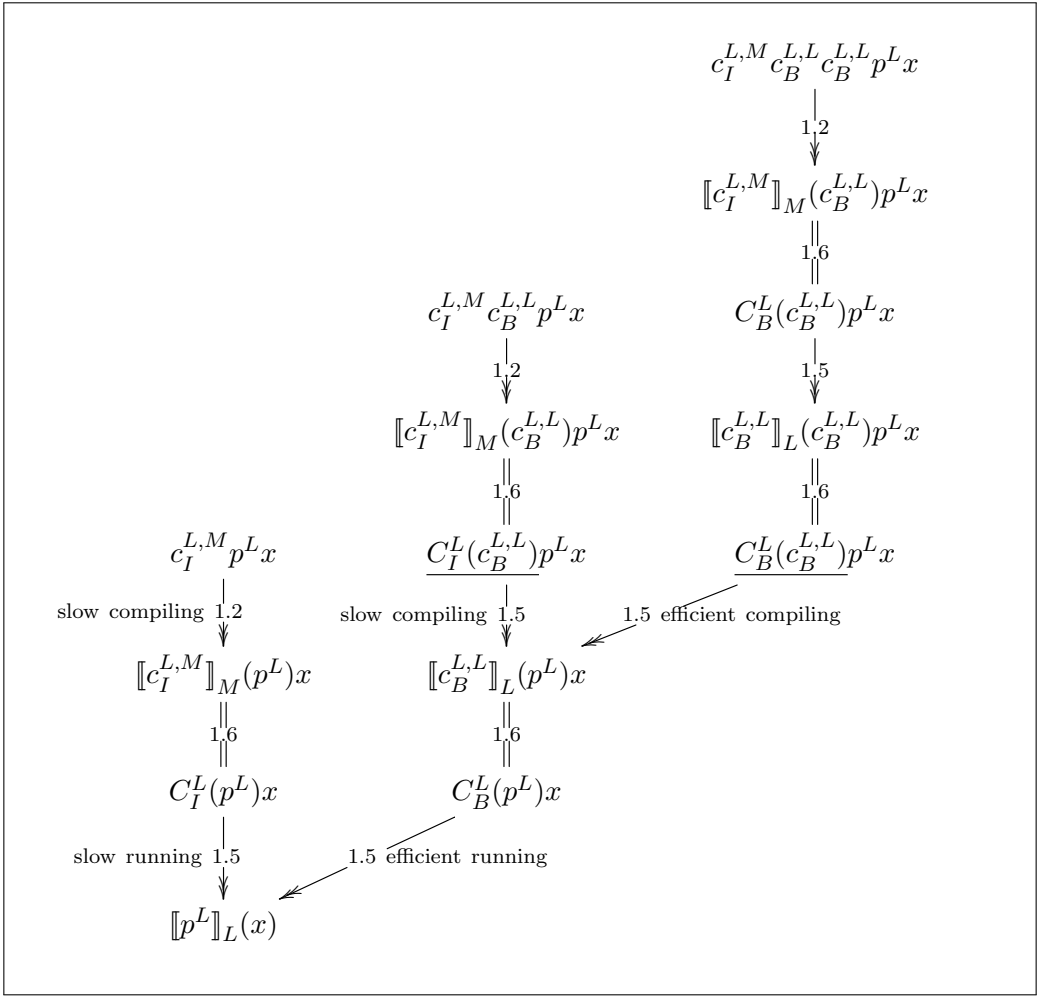
  }$$} \figcaption{Bootstrapping: precompiled
  $c_B^{L,M}:=C_I^L(c_B^{L,L})$, $c_{B'}^{L,M}:=C_B^L(c_B^{L,L})$
  provide efficient run time alone, or both run time and compile time,
  respectively.}\efig

\hoog{1.5em}In the language of combinatory logic, so much admired by Corrado
B\"ohm, one writes $p\cdot x$, or simply $px$, for $\set{p}_M(x)$, and
$cpx$ for $(cp)x$, etcetera (association to the left).  Then the three
ways of compiling and computing a job $\sem{p^L}_L(x)$ can be rendered
as in Figure 1.  The underlined expressions denote the codes of the
B\"ohm compiler $c_B^{L,L}$ that are obtained by precompilation,
respectively using the interpreter and using itself.  So in the steps
above these do not require time. This bootstrapping process wasn't
discussed in B\"ohm's PhD thesis, but it was made possible by his
invention and implementation of self-compilation. In Figure \ref{two}
the bootstrap process is presented in a slightly different way.

\vspace{1em}
\bfig
\hspace{-2.99em}\wit{0.000}\fbox{\btab{rl}
$\xymatrix{
  &x\ar[d]&\\
  \wit{10}p^M\ar@{->}[r]&
  \bullet\ar@{->}[r]&y=\sem{p^M}_M(x)
}$&
\raisebox{-3.8em}{\scriptsize\btab[b]{l}
The universal machine $\bullet$\\
with program $p^M$ and input $x$.\\
Although $M=L_M$ is a uni-\\
versal language, it is difficult to\\
write creative programs in it.
\etab}\\
$\xymatrix{
  &p^L\ar@{->}[d]&x\ar[d]&\\
  c_I\ar[r]&\bullet\ar@{->}[r]|-{p^M}&
  \bullet\ar@{->}[r]&y=\sem{p^L}_L(x)
}$&\raisebox{-3.5em}{\scriptsize\btab[b]{l}
  Using a simple compiler (inter-\\
  preter) $c_I\colon L\to M$ one can code\\
   better programs in $L$ to run\\
   on $M$. These have slow com-\\
   pile-time and slow run-time.
\etab}\\
$\xymatrix{
  &c_B^L\ar[d]&p^L\ar@{->}[d]&x\ar[d]&\\
  c_I\ar[r]&\bullet\ar[r]|-{c_B^M}&\bullet\ar@{->}[r]|-{p_*^M}&
  \bullet\ar@{->}[r]&y=\sem{p^L}_L(x)
}$&\raisebox{-2.5em}{\scriptsize\btab[b]{l}
  Using $c^L_B$, with compiling\\
  function yielding optimized\\
  code, one can obtain efficient\\
  run-time, but not compile-time.
\etab}\\
$\xymatrix{
  &c^L_B\ar[d]&c_B^L\ar[d]&p^L\ar@{->}[d]&x\ar[d]&\\
  c_I\ar[r]&\bullet\ar[r]|-{c^M_B}&
  \bullet\ar[r]|-{c_{B*}^M}&\bullet\ar@{->}[r]|-{p_*^M}&
  \bullet\ar@{->}[r]&y=\sem{p^L}_L(x)
}$&\raisebox{-3em}{\scriptsize\btab[b]{l}
  Compiling $c^L_B$ by itself yields \\
  an optimized $c_{B*}^M$ from which\\
  one can obtain efficient\\
  run-time and compile-time.
\etab}
\etab}
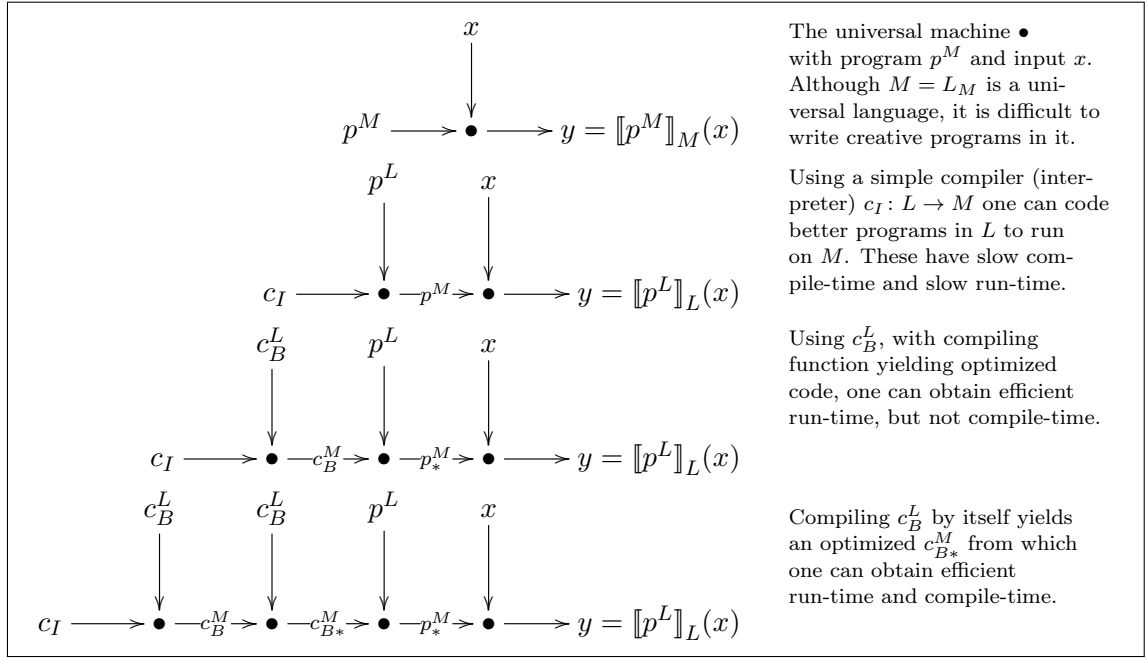
\figcaption{A different perspective on the same bootstrap process.}\label{two}
\efig

\hoog{1.6em} After having obtained his PhD in Z\"urich, B\"ohm did
succeed registering a patent on compilers.  But, unexpectedly, a few
years later (1955) IBM came with its FORTRAN compiler. It turned out
that B\"ohm's patent was valid only in Switzerland!

\subsection{Compiler configurations}
In this section we treat compilers in greater generality, translating
a language $L_1$ into $L_2$. Only one machine $M$ is used for the
translation, but this easily can be generalized. We settle the
question whether it is necessary to have self-compiling, in order to
make compile-time and run-time both efficient.

\bdf\bsub\fit We define the language  $\CCC$ of
\emph{compiler configurations} by the following context free grammar. 
$$\CCC::= L \mid (L,\bC_1,c,\bC_2),\mbox{ where $c\in L$ and
  $\bC_1,\bC_2\in\CCC$.}$$ Actually $L$ is a symbol $\ul{L}$ for a
language $L$, but we identify the two.
\item Let $\C\in\CCC$. The \emph{language} of $\C$,
  in notation $|\C|$, is defined as follows.  
\beqn
|L|&=&L;\\
|(L,\bC_1,c,\bC_2)|&=&L.
\eeqn 
\item Correctness of $\C\in\CCC$~~is defined as follows.\dwn \btab{rcl}
  $L$ &is correct;&\\ $(L,\bC_1,c,\bC_2)$&is correct&if 
$c$ is a program in programming language {$|\bC_2|$} 
,\\
&&$\bC_1,\, \bC_2$ are
  correct and\\ 
&& {$\sem{c}_{|\bC_2|}\colon L\arr |\bC_1|$ } 
is a compiling function. 
  \etab \esub \edf

\bex  The three situations in Subsection \ref{1.4} can be
described as compiler configurations. {We use $c_0$ and $c_B$ instead
  of $c_I^{L,M}$ and $c_B^{L,L}$, respectively.}  \beqn
\bC_1&=&(L,M,c_0,M).\\ \bC_2&=&(L,M,c_B,\bC_1)=(L,M,c_B,(L,M,c_0,M)).\\ \bC_3&=&(L,M,c_B,\bC_2)=(L,M,c_B,(L,M,c_B,(L,M,c_0,M))).
\eeqn \eex

\bdf
A compiler configuration $\C$ can be drawn as a labeled tree $T_\C$.
\beqn T_L&=&L;\\
T_{(L,\bC_1,c,\bC_2)}&=&
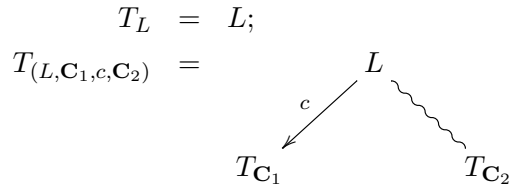
\xymatrix{&L\ar@{->}[dl]_c\ar@{~}[dr]&\\
T_{\bC_1}&&T_{\bC_2}}
\eeqn
\edf

Compiler configurations and their trees are more convenient to use
than the more rigid T-diagrams introduced in \cite{mcke70}, since
there is more flexibility to draw languages that still need to be
translated. For example, $\bC_3$ is the compiler configuration employed
by B\"ohm and its tree explains well the magic trick.

\bdf A compiler configuration $\C$ is inductively defined to be
\emph{executable} as follows.
$$\btab{ll}
$L$& is executable iff $L=M$;\\
$(L,\bC_1,c,\bC_2)$
& is executable iff $\bC_1$ and $\bC_2$ are executable.
\etab$$
\edf

\bex\bsub\fit
The three compiler configurations $\bC_1,\bC_2,\bC_3$ considered before
are executable.
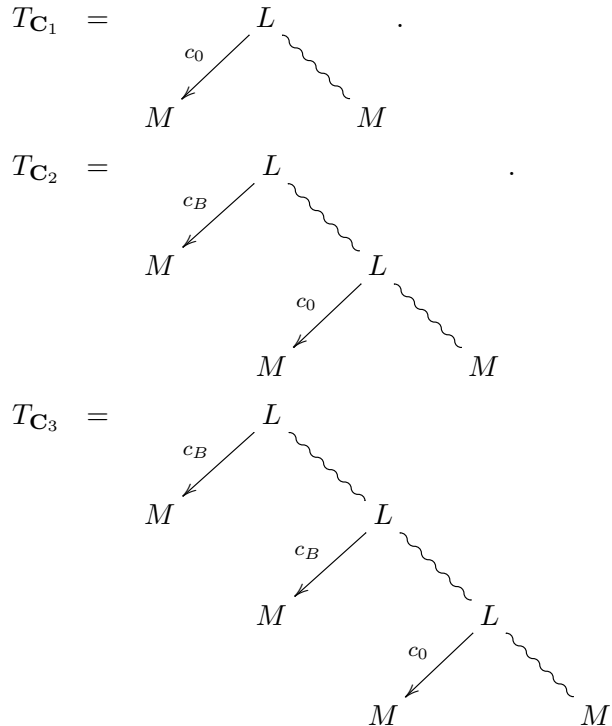
\beqn
T_{\bC_1}&=&
\xymatrix{&L\ar@{->}[dl]_{c_0}\ar@{~}[dr]&\\
M&&M}.\\
T_{\bC_2}&=&
{\xymatrix{&L\ar@{->}[dl]_{c_B}\ar@{~}[dr]&\\
M&&L\ar@{->}[dl]_{c_0}\ar@{~}[dr]\\
&M&&M}.}\\
T_{\bC_3}&=&
{\xymatrix{&L\ar@{->}[dl]_{c_B}\ar@{~}[dr]&\\
M&&L\ar@{->}[dl]_{c_B}\ar@{~}[dr]&\\
&M&&L\ar@{->}[dl]_{c_0}\ar@{~}[dr]\\
&&M&&M}.}
\eeqn
\item
The following compiler configurations, drawn as trees, are
not executable:
$$ T_\C= 
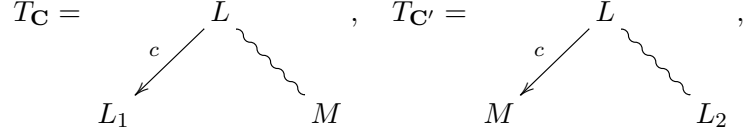
\xymatrix{&L\ar@{->}[dl]_c\ar@{~}[dr]&\\ L_1&&M},\quad
T_{\C'}= \xymatrix{&L\ar@{->}[dl]_c\ar@{~}[dr]&\\ M&&L_2},$$ because
an evaluation function for neither $L_1$ nor $L_2$ is given.
\esub\eex

\bdf To each $\C\in\CCC$~~we
assign a function that maps a program $p$ and value $x$ to a
value $\Phi_\C(p)(x)$, also written $\Phi_{\C}px$.
\beqn
\Phi_Lpx&=&\sem{p}_L(x);\\
\Phi_{L,\bC_1,c,\bC_2}px&=&\Phi_{\bC_1}(\Phi_{\bC_2}cp)x.
\eeqn \edf

\bexc
For all correct and executable $\C\in\CCC$, $p\in|\C|$, $x\in D$
one has
$$\Phi_\C px=\sem{p}_{|\C|}(x).$$
\eexc

\bex In the following
evaluations we leave out parenthesis, like in lambda calculus and
combinatory logic.
\[
\barr{rclll}
\Phi_Mp^Mx&=&\sem{p^M}_Mx&\dder\set{p^M}x&=p^Mx.\\
\Phi_{\bC_1}p^Lx&=&\sem{\sem{c_I}_Mp^L}_Mx&\dder \set{\set{c_I}p^L}x&=c_Ip^Lx.\\
\Phi_{\bC_2}p^Lx&=&\seM{\seM{\seM{c_I}c_B}p^L}x&\dder\set{\set{\set{c_I}c_B}p^L}x
&=c_Ic_Bp^Lx.\\
\Phi_{\bC_3}p^Lx&=&\seM{\seM{\seM{\seM{c_I}c_B}c_B}p^L}x
&\dder\set{\set{\set{\set{c_I}c_B}c_B}p^L}x&=c_Ic_Bc_Bp^Lx.  \earr\] \eex

Do we absolutely need self-compilation in order to obtain efficient
compilation?  The answer is negative.  Suppose one has the following:
\benum
\item a compiler $c_1^{L,L_1}: L \red M$, producing fast code, written in $L_1$;
\item a compiler $c_2^{L_1,L_2}: L_1 \red M$, producing fast code written in $L_2$;
\item a simple interpreter $c_I^{L_2,M}: L_2 \red M$, written in $M$.
\eenum
Then one can form the following correct and executable compiler configuration:
$$\bC_4=(L,M,c_1^{L,L_1},(L_1,L_2,c_2^{L_1,L_2},(L_2,M,c_I^{L_2,M},M))),$$
with tree
$$T_{\bC_4}=
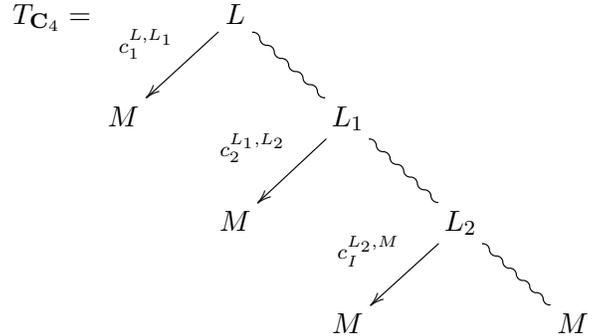
{\xymatrix{&L\ar@{->}[dl]_{c_1^{L,L_1}}\ar@{~}[dr]&\\
M&&L_1\ar@{->}[dl]_{c_2^{L_1,L_2}}\ar@{~}[dr]\\
&M&&L_2\ar@{->}[dl]_{c_I^{L_2,M}}\ar@{~}[dr]\\
&&M&&M}.}$$
Again one obtains a compiler with fast compile-time that
produces efficient code
$$c=C_2(c_1^{L,L_1})\dder\set{\set{c_I^{L_2,M}}(c_2^{L_1,L_2})}(c_1^{L,L_1}).$$
In the magic trick of B\"ohm, compiler (3) in Subsection \ref{1.4} above, 
he took $L=L_1=L_2$ and
$c_1^{L,L_1}=c_2^{L_1,L_2}=c_B^{L,L}$. This saves work: only one
language and one compiler need to be developed.

\section{Structured programming}
In a Turing machine transition a state can be followed by any other
state. Therefore many programming languages naturally contain the
`{\tt goto}' statement. When these are used in a mindless way, the
meaning of a program is not obvious, hence its correctness is much
more difficult to warrant. The first half of \cite{bohmjaco66} is
dedicated to eliminate goto statements, as a first step towards
structured programs. That part of the paper is stated to be written by
Jacopini, but I think we may suppose that B\"ohm, the supervisor of
Jacopini, has contributed to it.

\subsection{Imperative programming}
The Universal Turing Machine, or an improved version, immediately
gives rise to a language with {\tt goto} statements: the machine,
being in state $s_1$ changes (under the right conditions) into state
$s_2$.  This is expressed by a statement very much like a 5-tuple of a
Turing Machine $\lr{{\tt 1,s_1,0,s_2,N}}$, that in the presence of named
registers looks like
$${\tt s_1\colon\, if\; x= 1\; then\; x:=0;\; goto\; s_2;}$$ 
Here the meaning is as follows: the machine checks whether the content
of register $x$ equals 1 and then it overwrites the 1 by a 0 as the
content of register x, after which it jumps to state ${\tt s_2}$.  In
the presence of addressable registers like ${\tt x}$, there is no
longer a need to use the small step local movements indicated by
$\set{{\tt L,R,N}}$. A more extended example is the following.
$${\tt s_1\colon\, if\; x= 1\; then\; (y:=0;\; goto\; s_2)\; else\;
  (y:=y+1;\; goto\; s_3);}$$ Apart from branching, leading naturally
to a flow-chart as a representation of such a program, we also see the
for imperative programming typical statement ${\tt y:=y+1}$, meaning
that the content of register $y$ is overwritten by the old content
augmented by one. Many such components can form nice-looking but hard
to understand diagrams. One can imagine that the idea arose to create
more understandable diagrams and as a first step to eliminate the
{\tt goto} statements.

\subsection{Eliminating the `{\tt goto}'}

In this subsection it is shown that the result of eliminating the {\tt
  go to } statement can be seen in the light of Kleene's analysis of
computability, as was pointed out in \cite{coop67} and also in
\cite{hare80}.

\bthm[Kleene Normal Form Theorem]\label{KNFT} There are functions $U, T$
that are primitive computable such that
every computable function $f$ has a code
number $e$ such that for all $\vec{x}\in\nat$ one has 
$$f(\vec{x})=U(\mu z.T(e,\vec{x},z)=0).\eqno{\rm (NFT)}$$ If $P$ is a
predicate on $\nat$, then $\mu z.P(z)$ denotes the least number
$z\in\nat$ such that $P(z)$, if this $z$ exists, otherwise the
expression is undefined. In (NFT) it is assumed that for all $x$ there
exists a $z$ such that $T(e,x,z)$ holds\footnote{The formula (NFT)
  also holds for partial functions $f$, in which case $f(\vec{x})\up$
  iff $\forall z.T(e,\vec{x},z)\not=0$.}.  \ethm \bpf[Sketch] The value
of the function $f(\vec{x})=y$ can be computed by the Universal Turing
Machine $\cU$ using, say, $e$ as program. Then there is a computation
$${\tt
  (input,s_0,p_0)\rightarrow_{\cU}
  (t_1,s_1,p_1)\rightarrow_\cU\cdots\rightarrow_\cU(t_k,s_k,p_k)\rightarrow_\cU
  (output,s_h,p_h),}\eqno{\rm (comp)}$$ where ${\tt
  input}=(e,\vec{x})$, `${\tt input, s_0,p_0}$' is the first
configuration, `${\tt output, s_0,p_0}$' is the last one that is
terminating, and ${\tt output}=y$. Furthermore, ${T}$ is the
characteristic function ($=0$ when true, $=1$ when false) of the
primitive computable predicate $P(e,\vec{x},z)$, that holds if $z$ is
(the code of) the computation (comp). After a search (by $\mu$) for this
 (coded sequence) $z$, the $y={\tt output}$ is easily obtainable from
it, which is done by the primitive computable function $U$.\epf

\bthmC [\cite{bohmjaco66}]\label{BJ} A program built up
from statements of the form\dwn 
{\tt 
$\left.\btab{l}
x:=x+1\\ 
x:=x-1\\ 
if B, then $S_1$ else $S_2$\\
goto q 
\etab\right\}L_1$}\dwn can be
replaced by an equivalent one built up from statements of the form\dwn {\tt
$\hoog{1.9em}\left.\btab{l}  
x:=x+1\\ 
x:=x-1\\ 
if B, then $S_1$ else $S_2$\\
for k:=0 to n do A(k)\\ 
while x>0 do A(x)
\etab\right\}L_2$}\\
\ethmC \bpf[Sketch] A function $f$ with program from $L_1$
will be computable by the universal Turing Machine by
program, say, $e$. Therefore by Theorem \ref{KNFT} one has $f(\vec{x})=U(\mu
z.T(e,\vec{x},z)=0)$. The functions $U,T$ are primitive computable, hence
expressible by the `${\tt for}$' statements.  Only for the $\mu$ a
${\tt while}$ statement is needed. (Actually this happens only a
single time.)\epf \bcor[Folk Theorem]\label{CBJ}
Programs in $L_1$ can be replaced by an equivalent one in $L_2$ using 
the {\tt while} construct only a single time.
\ecor\bpf By the parenthetical remark in the proof of \ref{BJ}.\epf

\subsection{Evaluation}
After the {\tt goto} was shown to be eliminable, in Dijkstra's note
\cite{dijk68} a polemics was started `{\tt goto} statement considered
harmful'. In the book \cite{dahldijkhoar72} structured programming was
turned into an art. In \cite{knut74} it is argued that eliminating the
{\tt goto} as in the above proof of Theorem \ref{BJ} may produce
unstructured programs, unrelated to the original program.  The
original proof in \cite{bohmjaco66} does preserve the structure of the
program in a better way. See \cite{mill72} for a discussion.  An even
better way to eliminate the {\tt goto} statements, while preserving
the structure of a program, is described in \cite{ashcmann72}. An
example of a program in which a {\tt goto} statement does improve its
structure is also given in \cite{knut74}.

In \cite{hare80} the paper \cite{bohmjaco66} was taken as an example
of how a `Folk Theorem' appears.  The result attributed to these
authors often is Corollary \ref{CBJ}, rather than Theorem \ref{BJ}
itself.

As remarked in \cite{bohmjaco66} it seems necessary to use an extra
variable to obtain a program without a {\tt goto}, but the authors
couldn't find a proof of this conjecture.  A proof was given in
\cite{knutfloy71}, also in \cite{ashcmann72} and in
\cite{kozetsen08}.

Although the B\"ohm-Jacopini result started a discussion towards
structured programming, a new idea was needed to obtain even better
structured programs. As we will see in the next section, actually it
was an old idea: functional programming based on lambda calculus.

\section{Functional programming and the CUCH machine}
It was Wolf Gross, colleague of Corrado B\"ohm, who introduced the
latter to functional programming based on type-free lambda calculus,
in which unlimited self-application is possible. As can be imagined,
knowing the construction of a self-applicative compiler, it had a deep
impact on the sequel of B\"ohm's professional life. We restrict
ourselves and give some historical and conceptual background.

\subsection{Functional programming}
Alonzo Church introduced lambda calculus as a way to mathematically
characterize the intuitive notion of computability. I seem to remember
that he told me the following story. Church's thesis supervisor,
Oswald Veblen, gave him the problem to compute the Betti numbers of an
algebraic surface given by a polynomial equation. Church did not
succeed in this task and was stuck developing his PhD
thesis.  He then did what other mathematicians do in similar
circumstances: solve a different but related problem. Church wondered
what the notion `computable' actually means. Perhaps determining the
Betti number of a surface from its description is not a computable task.

Church then introduced a formal system for mathematical deduction and
computation \cite{chur32,chur33}. In \cite{kleeross35} his students
Kleene and Rosser found an inconsistency\footnote{The proof of a
  contradiction in Church's system was beautifully simplified in
  \cite{curr42}.} in Church's original system.  In \cite{chur36} the
system was stripped from the deductive part obtaining the (pure)
lambda calculus, which turned out to be provably consistent
\cite{churross36}. See \cite{bare84} for an extensive exposition of
the lambda calculus.

To formally define the notion of computability, Church introduced
numerals $\cnum{n}$ representing natural numbers $n$ as lambda terms.
Rosser found ways to add, multiply and exponentiate: that is, he found
terms $A_+,A_\times,A_{\sf exp}$ such that
$A_+\cnum{n}\cnum{m}\redd\cnum{n+m}$, and similarly for multiplication
and exponentiation. This way these three functions were seen to be
lambda definable. Here `$\redd$' denotes many-step rewriting, the
transitive reflexive closure of one-step rewriting `$\red$' introduced
below. At first neither Church nor his students could find a way to
lambda define the predecessor function.  At the dentist's office
Kleene did see how to simulate recursion by iteration and could in
that way construct a term lambda defining the predecessor function,
\cite{cros75}. (I believe Kleene told me it was under the influence of
laughing gas, ${\rm N_2O}$, used as anesthetic.) When Church saw that result he
stated ``Then all intuitively computable functions must be lambda
definable.'' That was the first formulation of Church's thesis and the
functional model of computation was born. At the same time Church gave
an example of a function that was non-computable in this model.

In \cite{turi37} it was proved that the imperative and functional
models of computation have the same power: they can compute exactly
the same partial functions, on say the natural numbers. The way these
computations are performed, however, differs considerably. In both
cases computations traverse a sequence of configurations, starting
essentially from the input leading to the output. But here the common
ground ends.

\subsection{Comparing imperative and functional programming}
In functional programming the argument(s) $A$ (or $\vec{A}\,$) for a
computation in the form of a function $F$ that has to be applied to
them form one single expression $FA$ (respectively $F\vec{A}\,$). Such
expressions are subject to rewriting. If the expression cannot be
rewritten any further, then the so called \emph{normal form} has been
reached and this is the intended output.  The intermediate results all
have the same meaning as the original expression and as the output. A
basic example of this is
$$(\lambda x.x^2+1)(3)\red 3^2+1\red 9+1\red 10,\eqno{(1)}$$ where $(\lambda x.x^2+2)$ is
the function $x\mapsto x^2+1$ that assigns to $x$ the value
$x^2+1$. In more complex expressions there is a choice of how to
rewrite, that is, which subexpression to choose as focus of attention
for elementary steps as above. For example not all choices will lead
to a normal form. There are \emph{reduction strategies} that always
will find a normal form if it exists. Normal forms, if they are
reached, are unique, the result is independent of choices how to
rewrite.  However performance, both time and space, is sensitive to the
steps employed.

In the imperative model a computation the configurations at each
moment of a computation sequence of a Turing Machine $M$ consist of
the momentaneous memory content on the \emph{tape}, the \emph{state}
of $M$, and \emph{position} of its head: ${\tt (t,s,p)}$. Each
terminating computation runs as follows:
$${\tt
  (input,s_0,p_0)\rightarrow_M(t_1,s_1,p_1)\rightarrow_M\cdots\rightarrow_M(t_k,s_k,p_k)\rightarrow_M
  (output,s_h,p_h),}\eqno{\rm (IP)}$$ where $\cour{s_h}$ is a halting
state (and ${\tt p_h}$ is irrelevant). The transitions ${\tt \rightarrow_M}$
depend on the set of instructions of the Turing Machine ${M}$. In the
case of non-termination the configurations never reach one with a
terminal state.  This description already shows that, wanting to
combine Turing Machines to form one that is performing a more complex
task, requires some choices of e.g.\ making the final state of the
first machine fit with the initial one of the second machine.

In the functional model of computation the sequence of configurations
is as follows:
$${\tt F\,input\redb E_1\redb \cdots \redb E_k\redb output.}\eqno{\rm
  (FP)}$$ All of these configurations are $\lambda$-terms and the
transitions $\redb$ are according to the single $\beta$-rule of
reduction, which is quite different. In order to make a more fair
comparison between the imperative and functional computation, one
could change (IP) and denote it as
$${\tt (input,c,s_0,p_0)\red[\cU](t_1,c,s_1,p_1)\red[\cU]\cdots\red[\cU](t_k,c,s_k,p_k)\red[\cU] (output,c,s_h,p_h)}\eqno{\rm (IP')},$$
where $c$ is the code (program) that makes the universal machine $\UU$
imitate the machine $M$. This makes (IP$'$) superficially similar to
(FP), that nevertheless is superior.

\subsection*{Advantages of functional programming}
In the sequence (FP) the expressions are
words in a language more complex than the simple strings in (IP) or
(IP$'$).

\benum\item[(i)] The $\lambda$-terms expressing functional programs have
the possibility of making abstraction upon abstraction, arbitrarily
often. This means that `components' of functions can be also
functions (of functions), enabling flexible procedures.
\item[(ii)] In FP there is no mention of state
and position, hence there is no need to deal with the bureaucracy of
these when combining programs. Hence FP has easy
compositionality.
\item[(iii)] In the sequence (FP) the meaning of each configuration
  remains the same, from the first to the last expression.  This can
  be seen clearly in the sequence (1) above.  \eenum Features (i) and
  (ii) of functional programs makes them transparent and compact.
  Feature (iii) makes it easier to prove them correct: reasoning with
  mathematical induction, substitution and abstraction often suffice;
  no need to learn new logical formalisms that are used to analyze
  imperative programs.  It can be expected that FP will become more
  and more important. The lack of side-effects makes it more easy to
  make parallel versions of programs.

\subsection*{Implementations of functional programming}
Functional Programming has been developed much more slowly than
Imperative Programming. The reason is that imperative programs can be
implemented rather directly on a Turing Machine or modern
computer. This is not the case for functional programs. Attempts to
construct specialized hardware for Functional Programming have not
been successful. But compilers from functional languages into ordinary
CPU's using imperative programs have been successfully developed.

One of the early examples is the SECD machine in \cite{land64}, soon
followed by work on the CUCH machine, \cite{bohmgros66},
\cite{bohm66}.  After fifty years of research on the use and
implementation of functional programming the field has come of
age. There exist fast compilers producing efficient code. One can
focus on the mathematical definition of the functions involved and the
correctness of these can be proved with relatively simple tools, like
substitution, abstraction and induction. A functional program is
automatically structured.  There are for example no `{\tt goto}'
statements.  See \cite{baremanzplas13} for a short description,
\cite{hugh89} for an extensive motivation, and \cite{peyt87} for
implementing functional programming languages.

\subsection*{Challenges for functional programming}
There are two main challenges for FP. 1.\ The lack of state makes
writing code for input/output more complex. 2.\ The evaluation result,
the output, doesn't depend on the way reduction takes place, but it is
not always easy to reason about space and time efficiency. These
issues are beyond the scope of this paper\footnote{Well known
  functional languages are LISP (later called Lisp), \cite{mcca62}
  (with many modern versions starting with Scheme \cite{scheme}), and
  ML \cite{milntoftharp97} (with modern version OCaml \cite{ocaml}).
  ML is loosely characterized as `Lisp with types' coming from the
  simply types lambda calculus, see \cite{chur40,curr34}, with a rich
  mathematical structure \cite{baredekkstat13}, Part I. However, Lisp
  and ML are not pure functional programming languages, in that they
  have assignment statements that can be used for input and output,
  making it also possible to write unstructured programs. In the pure
  functional languages, Haskell \cite{haskell} and Clean \cite{clean},
  at present the most developed ones, the I/O problem is solved by
  respectively monads and uniqueness typing. But using these features,
  in both cases it is still possible to write incomprehensible code
  when dealing with I/O.}.

\section{Separability in $\lambda$-calculus}
A mathematician is interested in numbers, not because these may
represent the amount of money in one's bank account (almost offensive
to mention), but for their properties definable from the basic
arithmetical operations $+$ and $\times$, such as primality.  Such a
love for numbers is not shared by most people. In the same way Corrado
B\"ohm became interested in $\lambda$-terms, not because they represent
programs that one can sell, but for their properties definable from
the basic lambda calculus operations: application and
abstraction. This is somewhat different from another form of
fascination, that of Donald Knuth for imperative programs that is obvious
from his volumes \cite{knut15}, driven by the challenge to
write clear, elegant, and efficient algorithms that perform relevant
computational tasks. We assume elementary knowledge of lambda calculus
and recall the following notations.

\bnot\bsub\fit The set of all lambda terms is denoted by $\Lambda$.
The set of free variables of $M\in\Lambda$ is denoted by $\FV{M}$.
The set $\Lo=\set{M\in\Lambda\mid \FV{M}=\emptyset}$ consists of the
\emph{closed} lambda terms without free variables, like $\lambda x.x,\lambda
xy.x$, but not $\lambda xy.z$.
\item `$\equiv$' denotes equality up to renaming bound
  variables, e.g.\ $\lambda x.x\equiv \lambda y.y$.
\item `=' denotes $\beta$-convertibility on $\lambda$-terms,
  often denoted by `$\eqb$' to be explicit.
\item $=_\beta$ is generated by $\beta$-reduction $\redb$, as in $(\lambda
  x.M)N\redb M[x:=N]$.
\item $=_\eta$ is generated by $\eta$-reduction
  $\rede$, as in $\lambda x.Mx\rede M$.
\item $M\in\Lambda$ is in \emph{$\beta(\eta)$ normal form} ($\beta(\eta)$-nf) if no $\redb$ (nor $\rede$) step is possible.
\item  For $\vect{M}{n}\in\Lambda$ write
  $\lr{\vect{M}{n}}\eqdf \lambda z.z\vecn{M}{n}$,
  with $z$ a fresh variable, i.e.\ $z\notin\FV{\vecn{M}{n}}$.
\item Write $\U^n_k\eqdf \lambda \vecn{x}{n}.x_k$.
  Note that $\lr{\vect{M}{n}}\U^n_k\eqb M_k$, for $1\leq k\leq n$.
\item Write\\[-2em]
\[  \bceqn
  \I&\eqdf &\lambda x.x;\\
  \K&\eqdf &\lambda xy.x,&\mbox{serving as `{\tt true}';}\\
  \K_*&\eqdf &\K\I =_\beta \lambda xy.y,&\mbox{serving as `{\tt false}';}\\
  \sS&\eqdf &\lambda xyz.xz(yz);\\
  {\sf C}&\eqdf &\lambda xyz.xzy;\\
  \Y&\eqdf &\lambda f.(\lambda x.f(xx))(\lambda x.f(xx));&\mbox{Curry's fixed point combinator;}\\
\Theta&\eqdf &(\lambda ab.b(aab))(\lambda ab.b(aab)),&\mbox{Turing's fixed point combinator;}\\
\om&\eqdf &\lambda x.xx;\\
\Om&\eqdf &\om\om, &\mbox{standard term without a nf};\\
\cnum{k}&\eqdf &\lambda fx.f^kx,&\mbox{where $f^0x=x$ and $f^{n+1}x=f(f^nx)$}\\
&&&\text{(Church's numerals)}.
  \eceqn\] \esub\enot

\subsection*{Separability of two normal forms}
\bdf
Terms $M_0,M_1 \in\Lo$ are called \emph{separable} if
for all $P_0,P_1\in\Lo$ there exists an $F\in\lambda$ such that 
$$FM_0\eqb P_0\und FM_1\eqb P_1.$$
This is equivalent to requiring that there is a lambda definable bijection
$$F\colon\set{M_0,M_1}/\!\eqb\;\;\red\;\set{\cnum{0},\cnum{1}}/\!\eqb,$$
with lambda definable inverse, in which case we write
$\set{M_0,M_1}=_1\set{\cnum{0},\cnum{1}}$.
\edf

In result \ref{B1} the principal step was proved in \cite{bohm68} with
the following result.

\bthmC[\cite{bohm68}]\label{BT} Let $M_0,M_1\in\Lo$ be two different
$\lambda$-terms in $\be$-nf. Then for all $P_0,P_1\in\Lo$ there exist
$\vec{N}\in\Lo$ such that \beqn
M_0\vec{N}&\eqb&P_0,\\ M_1\vec{N}&\eqb&P_1.  \eeqn \ethmC \bpf[Sketch]
A full proof (in English) is in \cite[Theorem 10.4.2]{bare84} and an
intuitive proof with applications in \cite{guerpipedeza09}.  Idea:
give the $M_0,M_1$ arguments separating the two. As we do not know in
advance which arguments will work, we use variables as unknowns and
substitute for them later. It suffices to reach two distinct variables, as
they can be replaced by $P_0,P_1$.  We present some
examples. 

\noi Example 1.\ $M_0\equiv\I,\,
M_1\equiv\K$. \\ ${\scriptsize \barr{l} \wit{3}xy\wit{14}
  {x{:=}\K\!\K}\wit{10}{zvw}\wit{30}{z{:=}\I}
  \earr}$\\ $\barr[c]{l|l|l|l|l}
\I&\I xy=xy&\K\K y=\K&\K zvw = zw& w\\
\K&\K xy=x& \K\K&\K\K zvw=\K vw=v& v \earr$
\wit{3}\raisebox{.6em}{\mbox{Hence  $\barr[t]{rcl} \I(\K\K)\I P_1P_0&=&P_0; \\
    \K(\K\K)\I P_1P_0&=&P_1.  \earr$}}\dwn
\begin{minipage}{\textwidth}
\noi Example 2.\ $M_0\equiv\I,\, M_1\equiv\om$.\hoogm\\
${\scriptsize \barr{l} \wit{3}x\wit{14}
  {x{:=}\K_*}\wit{24}{xyz}\wit{12}{x{:=}\K_*,y:=Ku}
  \earr}$\\ 
$\barr[l]{l|l|l|l|l}
\I&\I x=x&\K_*&\K_*xyz  = yz& x\\
\om&\om x=xx&\K_*\K_*=\K\I\K_*= \I&\I xyz=xyz& z
\earr$ \wit{3}\raisebox{.5em}{\mbox{Hence 
$\barr[t]{rcl}
\I\K_*\K_*(\K P_0)P_1&=&P_0;\\
\om\K_*\K_*(\K P_0)P_1&=&P_1.
\earr$}}\dwn
\end{minipage}
\begin{minipage}{\textwidth}
Example 3.\
$M_0\equiv\lambda xy.xy\I,\; M_1\equiv \lambda xy.xy\om$.
Consider these as trees:
$$
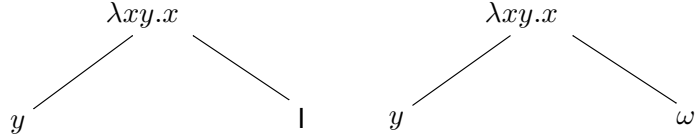
\xymatrix{&\lambda xy.x\wit{5}\\
y\ar@{-}[ru]&&\I\wit{5}\ar@{-}[lu]
}\quad
\xymatrix{&\lambda xy.x\wit{5}\\
y\ar@{-}[ru]&&\om\wit{3}\ar@{-}[lu]
}
$$
In order to separate these, we zoom in on the difference $\I$
and $\om$, via $M_0\K_*y, M_1\K_*y,$ giving $\I,\;\om$ respectively, 
and we know how to separate these by Example 2.\dwn
\end{minipage}
Example 4. $M_0\equiv\lambda xy.xy(x\I y),\; M_1\equiv \lambda xy.xy(x\om y)$.
Consider their trees:
$$
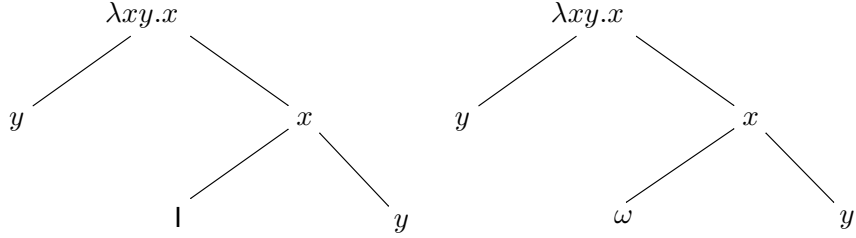
\xymatrix{&\lambda xy.x\wit{5}\\
y\ar@{-}[ru]&&x\ar@{-}[lu]\\
&\hspace{2em}\I\wit{3}\ar@{-}[ru]&&y\ar@{-}[lu]
}\quad
\xymatrix{&\lambda xy.x\wit{5}\\
y\ar@{-}[ru]&&x\ar@{-}[lu]\\
&\hspace{2em}\om\wit{3}\ar@{-}[ru]&&y\ar@{-}[lu]
}\\[-0em]$$
Again we like to zoom in on the difference $\I$ and $\om$. Dilemma:
one cannot make the $x$ choose both the left and right branch.
Solution: applying the `B\"ohm transformation'
\hb{$xy$, $x{:=}\lambda abz.zab$} postpones the choice and yields trees\\[-0em]
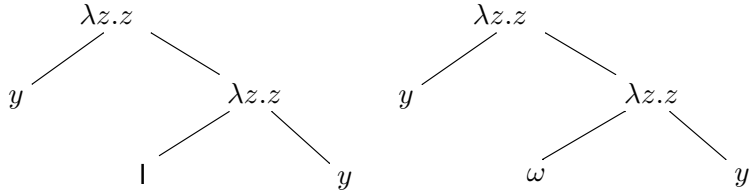
\[\xymatrix@-2ex{&\hb{\lambda z.z}\wit{5}\\ y\ar@{-}[ru]&&\hb{\lambda
    z.z}\ar@{-}[lu]\\ &\hspace{2em}\I\wit{3}\ar@{-}[ru]&&y\ar@{-}[lu]
}\quad \xymatrix@-2ex{&\hb{\lambda z.z}\wit{5}\\ y\ar@{-}[ru]&&\hb{\lambda
    z.z}\ar@{-}[lu]\\ &\hspace{2em}\om\wit{3}\ar@{-}[ru]&&y\ar@{-}[lu]
}\]
after which one can zoom in by application to
\hb{$z,z{:=}\K_*,z,z{:=\K}$}, obtaining $\I$ and $\om$; then we are back
to Example 2. Note that the dilemma was solved essentially by
replacing $x$ by $\lambda z.z$,
enabling to make postponed choices: first $\K$ (going right), then
$\K_*$ (going left).\epf

It is clear that one needs to require that the terms have different
$\be$-nfs, not just $\beta$-nfs. The terms $\lambda x.x$ and $ \lambda xy.xy$
are different $\beta$-nfs, but cannot be separated:
$F(\lambda x.x)\eqb \lambda xy.x$ and $F(\lambda xy.xy)\eqb \lambda xy.y$ would imply
$$\lambda xy.x\eqb F(\lambda x.x)=_\eta F(\lambda xy.xy)\eqb\lambda xy.y,$$ from which any
equation can be derived, contradicting that the $\lambda\be$-calculus is
consistent.

\bcor\label{B1}
For all $M_0,M_1\in\Lo$ having a
$\beta$-nf the following are equivalent.
\bsub
\item For all $P_0,P_1\in\Lo$ there exist $\vec{N}\in\Lo$
such that
$$M_0\vec{N}\eqb  P_0\und M_1\vec{N}\eqb  P_1.$$
\item $M_0,M_1$ are separable, i.e.\ for all 
 $P_0,P_1\in\Lo$ there exists an $F\in\Lo$
such that $$FM_0\eqb  P_0\und FM_1\eqb  P_1.$$
\item There exists an $F\in\Lo$ such that 
$$FM_0\eqb \lambda xy.x\und FM_1\eqb \lambda xy.y.$$
\item The equation $M_0=M_1$ is inconsistent with $\lambda\beta$.
\item The equation $M_0=M_1$ is inconsistent with
  $\lambda\be$\footnote{Dropping the requirement that both $M_0,M_1$
  have a $\beta$-nf, the result no longer holds: the equation
  $\lr{\K,\K_*}=\lr{\Omega\I,\Omega \cnum{1}}$ is consistent with
  $\lambda\beta$, but not with $\lambda\be$, as follows from considerations
  similar to those in \cite[Theorem 3.2.24]{barendregt20}.}.
\item The terms $M_0,M_1$ have distinct $\be$-nfs.  \esub
  \ecor\bpf{\rm \cite{statbare05}}

(1)$\Rightarrow$(2) 
By (1) there are $\vec{N}$ such that $M_i\vec{N}\eqb P_i$. Take
$F\eqdf\lambda m.m\vec{N}$.

(2)$\Rightarrow$(3) Take $P_i\eqdf\lambda x_0x_1.x_i$, for $0\leq i\leq 1$.

(3)$\Rightarrow$(4) From the equation $M_0=M_1$ one can by (3)
derive $\lambda xy.x=\lambda xy.y$, from which one can derive any equation; all
derivations using just $\lambda\beta$.

(4)$\Rightarrow$(5) Trivial.

(5)$\Rightarrow$(6) By the assumption that $M_0, M_1$ have $\beta$-nfs
and \cite{bare84}, Corollary 15.1.5, it follows that $M_0,M_1$ have
$\be$-nfs. If these were equal, then $M_0=_\be M_1$ and hence
$M_0=M_1$ would be consistent, contradicting (5).

(6)$\Rightarrow$(1) By Theorem \ref{BT}.
\epf

\subsection*{Separability of finite sets of normal forms}
Together with his students B\"ohm generalized Theorem \ref{BT} from
two to $k$ terms.

\bdf
A finite set $\cA\sbs\Lo$ is called \emph{separable} if for some $k\in\nat$
$$\cA=_1\set{\cnum{0},\ldots,\cnum{k-1}}.$$
\edf

\bthmC[\cite{bohmdezapereronch79}] Let $M_0,\ldots,M_{k-1}
\in\Lo$ be terms having different
$\be$-nfs. Then $\set{M_0,\ldots,M_{k-1}}$ is separable. One even has
for all terms $P_0,\ldots,P_{k-1}\in\Lo$ there exist
terms $\vec{N}\in\Lo$ such that
\beqn
M_0\vec{N}&\eqb&P_0,\\
&\ldots&\\
M_{k-1}\vec{N}&\eqb&P_{k-1}.
\eeqn\ethmC\bpf
For a proof see \cite{bohmdezapereronch79} or
\cite[proof of Corollary 10.4.14.]{bare84}.
\epf

\bcor Let $\cA\sbs\Lo$ be a finite set of terms all having a
$\beta$-nf. Then
\[\mbox{$\cA$ is separable $\iff$ the $\be$-nfs of the elements of $\cA$ are mutually different.}\]\qedhere
\ecor

\subs{Separability of finite sets of general terms}


A characterization of separability for finite $\cA\sbs\Lo$, possibly
containing terms without normal form, is due to \cite{coppdezaronc78},
see also \cite{bare84}, Theorem 10.4.13. To taste a flavor of that
theorem we give some of its consequences collected in
\cite{statbare05}.  \benum
\item The set
$\left\{\barr[c]{l}
\lambda x.x\cnum{0}\Om,\\
\lambda x.x\cnum{1}\Om
\earr\right\}$
is separable; so is
$\left\{\barr[c]{l}
\lambda xy.xx\Om,\\
\lambda xy.xy\Om
\earr\right\}$.
\item 
$\left\{\barr[c]{l}
\lambda x.x(\lambda y.y\Om),\\
\lambda x.x(\lambda y.y\cnum{0})
\earr\right\}$
is not separable; neither is
$\left\{\barr[c]{l}
\lambda x.x,\\
\lambda xy.xy
\earr\right\}$.
\item  
$\left\{\barr[c]{l}
\lambda x.x(\lambda y.y\cnum{0}\Om(\lambda z.z\Om)),\\
\lambda x.x(\lambda y.y\cnum{1}\Om(\lambda z.z\cnum{1})),\\
\lambda x.x(\lambda y.y\cnum{1}\Om(\lambda z.z\cnum{2}))
\earr\right\}$
is separable.
\item
$\left\{\barr[c]{l} 
\lambda x.x\cnum{0}\cnum{0}\Om,\\ 
\lambda x.x\cnum{1}\Om\cnum{1},\\ 
\lambda x.x\Om\cnum{2}\cnum{2} 
\earr\right\}$
is not separable, although each proper subset is. \eenum

\subs{Separability of infinite sets of general terms}

In \cite{statbare05} for infinite sets separability is defined and
characterized. Here we give a slightly alternative formulation.

\bnnot
Let $\cA\sbs\Lo$. Write for $F\in\Lo$
$$\barr{l}
F\cA\eqdf\set{FM\mid M\in\cA};\\
\CC_\nat\eqdf \set{\cnum{n}\mid n\in\nat}.
\earr$$
\ennot

\bdf\label{def.separability.gen}
Let $\cA\sbs\Lo$ be an infinite set. Then
\bsub\item  $\cA$ is called
\emph{special} if there are combinators $F,G\in\Lo$ such that
modulo $\eqb$ one has
\[\bceqn
F\colon \cA\red \bC_\nat&&&\mbox{is an injection},\\
G\colon \CC_\nat\red \cA&&&\mbox{is a surjection.}
\eceqn\]
\item
  $\cA$ is called \emph{separable} if $\cA=_1\CC_\nat$, that is,
  there is a lambda definable bijection
  $F\colon\cA\red\CC_\nat$ with lambda definable inverse.
  \esub
\edf

\brem If $\cA$ only has a $\lambda$-definable $F\colon \cA\red \CC_\nat$
injection, then $\cA$ doesn't need to be special. Indeed, let
$K\subseteq\nat$ be re but not recursive, so that its complement
$\overline{K}\sbs\nat$ is not re. Define $\cA=\set{\cnum{n}\mid
  n\in\overline{K}}$. Then $\I\colon\cA\red\CC_\nat$ is an injection.
For this $\cA$ there is no $\lambda$-definable surjection $G\colon\CC_\nat\red\cA$,
for otherwise
  \[\bceqn
  n\in\overline{K}&\iff&\cnum{n}\in\cA\\
  &\iff&\exists m.\cnum{n}\eqb G\cnum{m},&\mbox{which is re,}
  \eceqn\]
  contradicting that $\overline{K}$ is not re.
  \erem

  \bdf  $\cA$ is called an \emph{adequate numeral system} if there are terms
  $\ul{0},\ul{S},\ul{P},\ul{Z}_?$ (zero, successor, predecessor, test
  for zero) such that, writing $\ul{n}\eqdf\ul{S}^n\ul{0}$ for
  $n\in\nat$, one has \beqn \cA&=&\set{\ul{n}\mid
    n\in\nat};\\ \ul{P}(\ul{n+1})&=&\ul{n};\\ \ul{Z}_?\ul{0}&=&\lambda
  xy.x;\\ \ul{Z}_?\ul{n+1}&=&\lambda xy.y.  \eeqn 
\edf

\bprop\label{sep.imp.bij}
Let $\cA\sbs\Lo$ be infinite. If
$\cA$ is special, then there is a lambda definable bijection
$H\colon\cA\red\CC_\nat$.
\eprop
\bpf
Let combinators $F,\, G$ be given as required in Definition
\ref{def.separability.gen}. Define by primitive recursion
\[\bceqn
H\cnum{0}&=& G\cnum{0};\\
H\cnum{n+1}&=&
G{\cnum{\mu m.(G\cnum{m}\ninb\set{H\cnum{0},\ldots,H\cnum{n}})}},&(*)
\eceqn\]
In (*) `$\mu m$' stands for `the least number such that', which
in this case always exists
since $\cA$ is infinite and $G$ surjective. That $H$ is $\lambda$-definable
follows from the existence of $F$: indeed,
for $M,N\in\cA$ one has $$M\not=_\beta N\iff FM\not=_\beta FN\iff
\neg Q_=(FM)(FN),$$
where $Q_=$ is the decidable equality predicate on Church numerals,
so that also \[G\cnum{m}\ninb\set{H\cnum{0},\ldots,H\cnum{n}}\iff
\forall k\leq n.\neg Q_=(F\circ G\cnum{m})(F\circ H\cnum{k})\]
is decidable.

Claim. For all $n\in\nat$ one has
$$\set{G\cnum{0},\ldots,G\cnum{n}}\sbs
\set{H\cnum{0},\ldots,H\cnum{n}}.$$
The claim follows by induction on $n$.
Case $n=0$. By definition
$H\cnum{0}=G\cnum{0}$.

Case $n+1$. Assume $\set{G\cnum{0},\ldots,G\cnum{n}}\sbs
\set{H\cnum{0},\ldots,H\cnum{n}}$ (induction hypothesis), towards
$\set{G\cnum{0},\ldots,G\cnum{n+1}}\sbs
\set{H\cnum{0},\ldots,H\cnum{n+1}}$.  If $G\cnum{n+1}\in
\set{H\cnum{0},\ldots,H\cnum{n}}$, then we are done.  Otherwise
$G\cnum{n+1}{\notin} \set{H\cnum{0},\ldots,H\cnum{n}}$. For
$m{<}(n{+}1)$ one has $G\cnum{m}\in
\set{G\cnum{0},\ldots,G\cnum{n}}$ which is a subset of 
$\set{H\cnum{0},\ldots,H\cnum{n}}$ by the induction hypothesis. Therefore
by definition $H\cnum{n+1}=G\cnum{n+1}$, and the conclusion holds again.
This proves the claim.

By clause (*) in the definition above $H$ is injective.
That it is also surjective follows from the claim and the surjectivity
of $G$.
\epf

\bcorC[\cite{statbare05}]\label{char.inf.sep} Let $\cA\sbs\Lo$ be infinite. Then the following are equivalent.\\[-1em]
\bsub\item
$\cA$ is special.
\item $\cA$ is separable.
\item $\cA$ is an adequate numeral system.  \esub\ecorC\bpf (1)
  $\Right$ (2). If $\cA$ is separable, via $F\colon\cA\red\CC_\nat$ and
  $G\colon \CC_\nat\red\cA$, then by Proposition \ref{sep.imp.bij} there
  exists a $\lambda$-definable $H\colon \CC_\nat\red \cA$ that is a
  bijection.  We need to show that $H$ has a $\lambda$-definable inverse. This
  $H^{-1}\colon\cA \red \CC_\nat$ can be defined by
  \[\bceqn
  H^{-1}&=& \lambda a.(\mu m.Hm\isb a)\\
  &=& \lambda a.(\mu m. F(Hm)\isb Fa)\\
  &=& \lambda a. (\mu m. Q_=(F{\circ} H m)(Fa)),&\mbox{as in the proof of the proposition.}
\eceqn\]

  (2) $\Right$ (3). By $H, H^{-1}$ the set $\cA$ inherits the structure of an
  adequate numeral system from  $\CC_\nat$.

  (3) $\Right$ (1). Let $\ul{0},\ul{S},\ul{P},\ul{Z}_?$ give $\cA$ the
  structure of an adequate numeral system. Then the computable
  functions can be $\lambda$-defined w.r.t.\ the $\ul{n}$. By primitive
  recursion on the $\ul{n}$ and $\cnum{n}$ numerals, respectively, one
  can define $\lambda$-definable $F\colon\cA\red \CC_\nat$ and
  $G\colon\CC_\nat\red\cA$ satisfying
  $F\ul{n}=\cnum{n}$ and $G\cnum{n}=\ul{n}$, making $\cA$ separable.
  \epf
  
\section{Translating without parsing}
Combinatory terms, built-up from ${\bf K},{\bf S}$ with just
application, with reduction rules
\[{\bf K}PQ\red{w} P,\;{\bf S}PQR\red{w} PR(QR),\]
suffice to represent arbitrary computations. We write all
parenthesis. For example ${\bf ((S(KK))S)}$ is such a term.  It was
noticed by B\"ohm and Dezani
that the meaning of such a term can be found by interpreting it symbol
by symbol, including the two parentheses. One doesn't need to parse
the combinator to display its tree-like structure. The method also
applies to combinatory terms build from different combinators,
including for example ${\bf B}$ corresponding to the $\lambda$-term $\B=\lambda
fgx.f(gx)=\lambda fg.f\circ g$.

\bdf Define for $\lambda$-terms $M,N$
\beqn M\circ N&=&\lambda x.M(Nx);\\
M\ddot N&=&N\circ M;\\
\lr {M}&=&\lambda x.xM.
\eeqn\edf
It is easy to see that $\circ$ and $\ddot$ are associative modulo
$\beta$-equality of the $\lambda$-calculus; moreover, for $k\geq 2$ one has
$$M_k\circ\ldots\circ M_1\circ M_1=\lambda x.M_k(\ldots(M_1(M_1 x))..).$$

\bdf Combinatory terms $\CC$ are built up over alphabet
$\Sigma=\set{\bK,\bS,(,)}$ by the following context-free grammar
$$\CC::=\bK\mid \bS \mid (\CC\,\CC)$$ \edf \bdf Given $P\in\CC$ its
translation into closed terms of the $\lambda$-calculus is $P_{\lambda}$ defined
recursively as follows: \beqn
\bK_{\lambda}&=&\K=\lambda xy.x;\\ 
\bS_{\lambda}&=&\sS=\lambda xyz.xz(yz);\\ 
(QR)_{\lambda}&=&Q_{\lambda} R_{\lambda}.  \eeqn \edf
For this translation the $P\in\CC$ needs to be parsed. For example if
$P=(QR)$, we need to know where the string $Q$ ends and similarly
where $R$ starts. The following translation avoids
this need for parsing.

\bdf\bsub\fit The symbols of $\Sigma$ are translated into $\Lo$ as follows.
\[\barr{lcc}
\#\,(&=&\B\\
\#\,\bK&=&\lr{\K}\\
\#\,\bS&=&\lr{\sS}\\
\#\;)&=&\I \earr\]
\item A word in $w=\vecn{a}{n}\in\Sigma^*$ is translated into
  $\phi(w)\in\Lo$ as follows.
$$\phi(w)= \#a_1\ddot\cdots\ddot\#a_n.$$
\esub\edf

\bpropC[\cite{bohmdeza72}]\label{BD}\bsub\fit
For all $P\in\CC$ one has $\phi(P)=_\beta\lr{P_{\lambda}}$.
\item
For all $P\in\CC$ one has $\phi(P)\I=_\beta P_{\lambda}$.
\esub\epropC\bpf\bsub\fit
Since $P\in\CC$, we may use induction over terms in $\CC$.
If $P=\bK$ or $P=\bS$, the result holds by definition of $\phi$.
If $P=(QR)$, then
\[\bceqn
\phi(P)&\isb&\#(\ddot\phi(Q)\ddot\phi(R)\ddot\#),&\mbox{ by the associativity of $\ddot$,}\\
&\isb&\B\ddot\phi(Q)\ddot\phi(R)\ddot\I,\\
&\isb&\I\circ\lr{Q_{\lambda}}\circ\lr{R_{\lambda}}\circ\B,&\mbox{ by definition of $\ddot$ 
and the ind.\ hyp.},\\
&\isb&\lambda x.\I(\lr{Q_{\lambda}}(\lr{R_{\lambda}}(\B x))),\\
&\isb&\lambda x.(\lr{Q_{\lambda}}(\B xR_{\lambda})),\\
&\isb&\lambda x.\B xR_{\lambda} Q_{\lambda},\\
&\isb&\lambda x.x(R_{\lambda} Q_{\lambda})=\lr {(RQ)_{\lambda}}=\lr{P_{\lambda}}.
\eceqn\]
\item By (1): $\phi(P)\I\eqb\lr{P_{\lambda}}\I\eqb \I P_{\lambda}\eqb P_{\lambda}.$\qedhere
  \esub\epf  Proposition \ref{BD}(2) shows that the
  meaning of $P$ can be obtained without parsing.

\section{A simple self-evaluator}
To $M\in\Lambda$ one assigns computably a G\"odel-number $\#M$.

\bdf
For $M\in\Lambda$ its code $\num{M}$
is defined as the Church numeral corresponding to $\#M$
$$\num{M}\eqdf\cnum{\#M}.$$ \edf Note that the code of $M$ satisfies
1.\ $\num{M}$ is in normal form; 2.\ syntactic operations on $M$ are
lambda definable on $\num{M}$, by the computability of $\#$.  An
evaluator $\sE$ is constructed by Stephen Cole Kleene in \cite{klee35}
such that for all $M\in\Lo$ one has
$$\sE\num{M}\isb M.$$ A technical problem to define $\sE$ and show this
is caused by the fact that the lambda terms are inductively defined
via open terms containing free variables. But the decoding only holds
for closed terms.  The way Kleene dealt with this (basically the
problem of representing the binding effect of $\lambda x$), was to
translate closed $\lambda$-terms first to combinators and then representing
these as numerals. The term $\sE$ was reconstructed by McCarthy for the
programming language LISP under the name `{\tt eval}',
and baptized in \cite{reyn72} as the `meta-circular'
self-interpreter.

During lectures at Radboud University on Kleene's self-evaluator $\sE$
and constructing this term via the combinators, the student Peter de
Bruin came with an improvement. He suggested to use the intuition of
denotational semantics of $\lambda$-calculus. First the meaning of an open
term $M$ (containing possibly free variables) is given, in notation
$\eval \num{M}v$, using a valuation $v$ assigning values
$v(\num{x})\in\Lambda$ to the code of a free variable $x$.

\bthmC[\cite{klee35}] There is a term $\sE\in\Lo$ such that
$$\forall M\in\Lo.\sE\num{M}\eqb M.$$ \ethmC
\bpf(P. de Bruin) By the
effectiveness of the G\"odel-numbering there exists an $\eval\in\Lo$
satisfying \beqn \eval \num{x} v&=&v(\num{x});\\ \eval
\num{(PQ)}v&=&(\eval \num{P}v)(\eval \num{Q}v);\\ \eval \num{(\lambda
  x.P)}v&=&\lambda y.\sE_0\num{P}(v[\num{x}\mapsto y]), \eeqn where
$v[\num{x}\mapsto y]=v'$ with
\[\bceqn
v'\num{z}&=&v\num{x},&\mbox{if $\num{z}\not=\num{x}$,}\\
v'\num{z}&=&y,&\mbox{if $\num{z}=\num{x}$.}
\eceqn\] Then one can prove that for $M\in\Lambda$ with
$\FV{M}\sbs\set{\vect{x}{n}}$ one has
$$E_0\num{M}v=M[\vect{x}{n}:=v(\num{x_1}),\ldots,v(\num{x_n})].$$
Therefore $$\forall M\in\Lo.\sE_0\num{M}v=M$$ and one can take $\sE\eqdf\lambda
m.\sE_0m\I$.\epf

\bcor
The term $\sE$ enumerates the closed $\lambda$-terms
$$\forall M\in \Lo\exists  n\in\nat.\sE\cnum{n}=M.$$
\ecor

\brem In \cite{bare95} it is proved (constructively) that any
enumerator of the closed terms is reducing in the following sense.
\[\forall M\in \Lo\exists  n\in\nat.\sE'\cnum{n}= M\imp
\forall M\in \Lo\exists  n\in\nat.\sE'\cnum{n}\redd M.\]
\erem

The construction of Peter de Bruin inspired \cite{moge92} to a higher
order encoding of $\lambda$-terms, see \cite{pfenelli88}, in which a $\lambda$
is interpreted by itself.

\bdfC[\cite{moge92}]
An open lambda term $M$ can be interpreted as an open lambda term
with the same free variables as follows.
\beqn \numm{x}&\eqdf &\lambda abc.ax;\\ \numm{PQ}&\eqdf &\lambda
abc.b\numm{P}\numm{Q};\\ \numm{\lambda x.P}&\eqdf &\lambda abc.c(\lambda x.\numm{P}).
\eeqn \edfC This can be seen as first using three unspecified
constructors ${\tt var, app, abs}\in\Lo$ as follows
\beqn
\numn{x}&\eqdf &\var\,x;\\ \numn{PQ}&\eqdf &\app\,\numn{P}\numn{Q};\\ \numn{\lambda
  x.P}&\eqdf &\abs\,(\lambda x.\numn{P}), \eeqn and then taking \beqn \var&\eqdf &\lambda
x\lambda abc.ax;\\ \app&\eqdf &\lambda pq\lambda abc.bpq;\\ \abs&\eqdf &\lambda
z\lambda abc.cz.  \eeqn

\bthmC[\cite{moge92}]\label{moge92}
There is an evaluator $\sE^m\in\Lo$ such that for all  $M\in\Lambda$
$$\sE^m\numm{M}\eqb M.$$ \ethmC\bpf Using Turing's fixed point combinator
$\Theta$ one can construct a term $\sE^m\in\Lo$ such that
$$\sE^mM\redd M\I (B\sE^m)(C\sE^m),$$
where $B\eqdf \lambda epq.ep(eq)$, and $C\eqdf \lambda ezx.e(zx)$: take
$\sE^m\eqdf \Theta(\lambda em.m\I(Be)(Ce))$. Then by induction on the structure of
$M\in\Lambda$ it follows that $\sE^m\numm{M}\redd M$.
\[\bceqn
\sE^m\numm{x}&\redd&\numm{x}\I(B\sE^m)(C\sE^m)\\
&\redd&\I x\red x;\\
\sE^m\numm{PQ}&\redd&\numm{PQ}\I(B\sE^m)(C\sE^m)\\
&\redd&B\sE^m\numm{P}\numm{Q}\\
&\redd&\sE^m\numm{P}(\sE^m\numm{Q})\\
&\redd&PQ,&\mbox{by the induction hypothesis;}\\
\sE^m\numm{\lambda x.P}&\redd&\numm{\lambda x.P}\I(B\sE^m)(C\sE^m)\\
&\redd&C\sE^m(\lambda x.\numm{P})\\
&\redd&\lambda x.\sE^m((\lambda x.\numm{P})x)\\
&\red&\lambda x.\sE^m\numm{P}\\
&\redd&\lambda x.P,&\mbox{by the induction hypothesis.}
\eceqn\]
\par \vspace{-1.3\baselineskip}
\qedhere
\epf

\brem\bsub\fit
Using Mogensen's translation, decoding is possible for all terms
$M\in\Lambda$ possibly containing free variables.  On the other hand
not all syntactic operations are possible on the coded terms.
Equality test for variables is possible for $\num{x}$, but not for
$\numm{x}$.  
\item 
  In spite of this, the lambda definability of equality discrimination
  for coded closed terms $\numm{M},\numm{N}\in\Lo$ is proved in
  \cite{bare01}.

\item In \cite{moge92} it is also proved that there is a
  normalizer acting on coded terms.
  
  \emph{There is a term $\R^m$ such that for all $M\in\Lambda$\\
\wit{15} if $M$ has a normal form $N$, then $\R^m\numm{M}\redd
    \numm{N}$;\\
\wit{15}  if $M$ has a no normal form, then $\R^m\numm{M}$ has no nf.
  } \esub \erem

  Berarducci and B\"ohm constructed a very simple self-evaluator,
  based on Mogensen's construction above, but using different choices
  for \var, \app, \abs. These are based on unpublished work of B\"ohm
  and Piperno, who represented algebraic data structures in such a way
  that primitive recursive (computable) functions are representable by
  terms in normal form, avoiding the fixed point operator that was
  used in the proof of Theorem \ref{moge92}.  \renewcommand\C{{\sf C}}

\bthmC[\cite{berabohm93}] There is a coding of $\lambda$-terms $M\mapsto\numb{M}$
with a short closed normal form $\Eb\eqdf \lr{\lr{\K,\sS,\C}}$ as evaluator.\ethmC\bpf
Define
\beqn
\numb{x}&\eqdf &\varb\,x;\\ \numb{PQ}&\eqdf &\appb\,\numb{P}\numb{Q};\\ \numb{\lambda
  x.P}&\eqdf &\absb\,(\lambda x.\numb{P}), \eeqn
where
\beqn \varb&\eqdf &\lambda x\lambda e.e\U^3_1xe;\\
\appb&\eqdf &\lambda pq\lambda e.e\U^3_2pqe\\
\absb&\eqdf &\lambda z\lambda e.e\U^3_3ze.\eeqn
By induction on the structure of $M$ we show that
$\numb{M}\lr{\K,\sS,\C}\redd M$.

Case $M\equiv x$. Then
\[\bceqn
\numb{x}\lr{\K,\sS,\C}&\redd&((\lambda x\lambda e.e\U^3_1xe)x)\lr{\K,\sS,\C}\\
&\redd&(\lambda e.e\U^3_1xe)\lr{\K,\sS,\C}\\
&\redd&\lr{\K,\sS,\C}\U^3_1x\lr{\K,\sS,\C}\\
&\redd&\K x\lr{\K,\sS,\C}\\
&\redd& x.
\eceqn\]

Case $M\equiv PQ$. Then
\[\bceqn
\numb{PQ}\lr{\K,\sS,\C}&\equiv&(\lambda pqe.e\U^3_2pqe)\numb{P}\numb{Q}\lr{\K,\sS,\C}\\
&\redd&\lr{\K,\sS,\C}\U^3_2\numb{P}\numb{Q}\lr{\K,\sS,\C}\\
&\redd&\sS \numb{P}\numb{Q}\lr{\K,\sS,\C}\\
&\redd& \numb{P}\lr{\K,\sS,\C}(\numb{Q}\lr{\K,\sS,\C})\\
&\redd&PQ,&\mbox{by the induction hypothesis.}
\eceqn\]

Case $M\equiv \lambda x.P$. Then
\[\bceqn
\numb{\lambda x.P}\lr{\K,\sS,\C} &\equiv&(\lambda ze.e\U^3_3ze)(\lambda
x.\numb{P})\lr{\K,\sS,\C}\\ &\redd&\lr{\K,\sS,\C}\U^3_3(\lambda
x.\numb{P})\lr{\K,\sS,\C}\\ &\redd&\C(\lambda
x.\numb{P})\lr{\K,\sS,\C}\\ &\equiv&(\lambda xyz.xzy)(\lambda
x.\numb{P})\lr{\K,\sS,\C}\\ &\redd&\lambda z.(\lambda
x.\numb{P})z\lr{\K,\sS,\C}\\ &\equiv&\lambda x.(\lambda
x.\numb{P})x\lr{\K,\sS,\C}\\ &\red&\lambda
x.\numb{P}\lr{\K,\sS,\C}\\ &\redd&\lambda x.P,&\mbox{by the induction
  hypothesis.}  \eceqn\] Therefore for all $M\in\Lambda$ one has
$\numb{M}\lr{\K,\sS,\C}\redd M$.  It follows that
$\Eb\eqdf\lr{\lr{\K,\sS,\C}}$ is a self-evaluator: for all $M\in\Lambda$
\[\Eb\numb{M}\equiv \lr{\lr{\K,\sS,\C}}\numb{M}
\red\numb{M}\lr{\K,\sS,\C}\redd M.\qedhere\]\epf It is a remarkable
coincidence that the term $\Eb\equiv\lr{\lr{\K,\sS,\C}}$ abbreviates
the name ``Kleene, Stephen Cole'', the full name of the inventor of
self-evaluation in $\lambda$-calculus. Corrado B\"ohm was fond of such
tricks and had for this and other reasons the nickname `il miracolo'.

\section*{Coda}
At a symposium in honor of Corrado B\"ohm's ninety's birthday, January
2013, at Sapienza University, Rome, the jubilee treated the audience with an
open problem. Actually it is more a `Koan' (not precisely stated) than a
Problem (with a precisely stated space of answers). But Koans
are often the more interesting problems in mathematics and computer science.\\

\begin{pkC}
  [(C.\ B\"ohm, 2013)]  Given $\beta$-normal
forms $F\equiv\lambda\vecn{x}{n}.P$, and $G\equiv\lambda \vecn{x}{n}.Q\in\Lo$.
By writing $F^d\eqdf\lambda x.F(x\cnum{1})\ldots(x\cnum{n})$ and similarly
for $G^d$, these terms can be made unary.  Trying to find closed terms
${M}$ 
from solutions $N$ of the equation $F^dN\eqb G^dN$? (Define a
\emph{deed} to be a closed nf of the form $\lambda x.x\vecn{P}{k}$. The
$F^d,G^d$ are deeds up to $\isb$.)
\end{pkC}

\section*{Acknowledgments}
The author thanks Marko van Eekelen for explaining him many years ago
the method of bootstrapping (Section 1), Mariangiola Dezani for
comments on the paper, and Rinus Plasmeijer for discussions about
Section 3. The referees provided very useful remarks, improving the
paper.

To the family of Corrado B\"ohm I am grateful for letting me spend
wonderful times with them, besides for fully enabling us to enjoy the
combinators.

\bibliographystyle{alpha}
\bibliography{CBML}
\end{document}